\documentclass[12pt]{article}
\usepackage{latexsym}
\usepackage{epsfig,amssymb,euscript,slashed}
\usepackage[linktocpage]{hyperref}
\usepackage{amsmath}
\usepackage{cite} 
\usepackage{array,calc,epsfig}
\usepackage{bbm}
\usepackage{fancybox}

\usepackage{comment}

\usepackage{tikz}
\usepackage{lmodern}
\usetikzlibrary{decorations.pathmorphing}
\usetikzlibrary{patterns}

\tikzset{
    photon/.style={decorate, decoration={snake}},
    electron/.style={postaction={decorate},
        decoration={markings,mark=at position .55 with {\arrow[draw=blue]{>}}}},
    gluon/.style={decorate,
        decoration={coil,amplitude=4pt, segment length=5pt}} 
}

\numberwithin{equation}{section} 
\usepackage[]{graphicx}
\usepackage{color}

\newcommand{\be}{\begin{equation}}
\newcommand{\ee}{\end{equation}}
\newcommand{\beq}{\begin{equation}}

\newcommand{\eeq}{\end{equation}}

\newcommand{\bra}{\langle}
\newcommand{\ket}{\rangle}

\newcommand{\half}{\frac{1}{2}}

\newcommand{\bs}{\begin{split}}
\newcommand{\es}{\end{split}}

\newcommand{\D}{\mathcal{D}}

\newcommand{\F}{\mathcal{F}}

\newcommand{\pd}{\partial}

\newcommand{\ep}{\varepsilon}

\begin{document}
\font\cmss=cmss10 \font\cmsss=cmss10 at 7pt

\begin{flushright}{
}
\end{flushright}
\hfill
\vspace{18pt}
\begin{center}
{\Large 
\textbf{AdS$_3$ four-point functions from $\frac{1}{8}$-BPS states}
}

\end{center}

\vspace{8pt}
\begin{center}
{\textsl{Alessandro Bombini$^{\,a,b,c}$, Andrea Galliani$^{\,c}$}}

\vspace{1cm}

\textit{\small ${}^a$ Dipartimento di Fisica ed Astronomia ``Galileo Galilei",  Universit\`a di Padova,\\Via Marzolo 8, 35131 Padova, Italy} \\  \vspace{6pt}

\textit{\small ${}^b$ I.N.F.N. Sezione di Padova,
Via Marzolo 8, 35131 Padova, Italy}\\
\vspace{6pt}

\textit{\small ${}^c$ Institut de Physique Th\'eorique, \\
Universit\'e Paris Saclay, CEA, CNRS, \\
Orme des Merisiers, F-91191 Gif sur Yvette, France}\\ 

\vspace{6pt}

\end{center}

\vspace{12pt}

\begin{center}
\textbf{Abstract}
\end{center}

\vspace{4pt} {\small
\noindent 
We compute four-point functions in the Heavy-Heavy-Light-Light limit involving a large family of $\frac{1}{8}$-BPS heavy states whose dual supergravity solutions are explicitly known, avoiding the use of Witten diagrams. This is achieved by using the AdS/CFT dictionary of type IIB supergravity on AdS$_3 \times S^3 \times {\cal M}_4$ that maps supersymmetric heavy operators whose conformal dimension is the order of the central charge to explicit asymptotically AdS supergravity solutions. 
Using the Ward Identities for the generators of the ${\cal N}=(4,4)$ superconformal $SU(2)$ Kac-Moody algebra, we can relate all of these four-point functions to each other and to other known four-point functions involving $\frac{1}{4}$-BPS heavy states, furnishing non-trivial checks of the computations. 
Finally, the Ward Identities can be employed to reconstruct the all-light four-point functions, providing the first holographic correlators of single-trace operators computed in AdS$_3$ involving $\frac{1}{8}$-BPS operators. 
}

\vspace{1cm}

\thispagestyle{empty}

\vfill
\vskip 5.mm
\hrule width 5.cm
\vskip 2.mm
{
\noindent  {\scriptsize e-mails:  {\tt alessandro.bombini@pd.infn.it, andrea.galliani@ipht.fr} }
}

\setcounter{footnote}{0}
\setcounter{page}{0}

\newpage


\tableofcontents

\newpage
\section{Introduction}
\label{sec:intro}

The study of correlators has been one of the key ingredients in AdS/CFT correspondence \cite{Witten:1998qj, Liu:1998ty, Freedman:1998bj, DHoker:1998ecp, DHoker:1999mqo} and, more recently, also in the context of black hole physics \cite{Lunin:2001jy, Balasubramanian:2005mg, Avery:2010qw}. In AdS/CFT context, black hole solutions are regarded as supergravity duals to a statistical ensemble of heavy states of the dual field theory, and the study of correlation functions provides a powerful tool to shed light on black hole physics and its puzzles. Among all the dynamical quantities one may focus on, four-point functions are especially relevant because of their nature as probes for the black hole and its microscopic structure. 

One of our main motivations for studying four-point functions is to confront the correlators computed in pure states, especially black hole \textit{microstates}, with those computed in a thermal background naively described, in the gravity side, by a black hole geometry. Correlators computed semi-classically in a black hole background share indeed properties that are in contrast with unitarity of the dual field theory. As pointed out firstly in \cite{Maldacena:2001kr}, one of the main subtle points arises looking at the late Lorentzian-time behaviour, that shows a decay \cite{Fitzpatrick:2016ive}. Indeed, regardless of the  pure or thermal nature of the states we are considering in the dual field theory, correlators have to satisfy a periodic behaviour at late-time and they cannot decay as they do in the naive picture \cite{Barbon:2003aq, Kleban:2004rx, Barbon:2004ce}. On the other hand, the general expectation for a consistent proposal describing microstates of a given black hole in gravity is to find correlators allowed by unitarity. All these tensions are in general considered as a different version of what is commonly called black hole {\it information paradox}. One of the technical aspects of this study is that, as often happens in holography, the regime where the semi-classical gravity description is reliable is usually dual to a strongly coupled field theory, a regime that is not treatable with standard field theory methods, and the necessity of a holographic way to extract quantities emerges.

A prototypical case in microstates physics is that of the Strominger-Vafa \cite{Strominger:1996sh} black hole, which admits an AdS$_3$ decoupling limit and a dual description in terms of a 2-dimensional SCFT, often dubbed as D1D5 CFT \cite{Seiberg:1999xz, Larsen:1999uk}. More in details, this CFT has ${\rm Vir}\otimes\overline{\rm Vir}$ symmetry, $\mathcal{N}=(4,4)$ supersymmetries and an $SU(2)_L\times SU(2)_R$ R-symmetry algebra. Its gravity dual is type IIB string theory on AdS$_3\times S^3\times\mathcal{M}_4$, where the four-dimensional compact space $\mathcal{M}_4$ can be either $T^4$ or K3 \cite{David:2002wn}. These background solutions arise as the near-horizon limit of a stack on $n_1$ D1 branes wrapping the $S^1$, and $n_5$ D5 branes wrapped on the common $S^1$ and on $\mathcal{M}_4$. The D1D5 CFT has a central charge $c=6N=6n_1n_5$ and its moduli space contains a free point where the CFT has a description in terms of a non-linear sigma model with target space the symmetrized orbifold  $\mathcal{M}_4^N/S_N$. This theory splits into two sectors for each chirality: a Neveu-Schwarz and a Ramond sector, connected by a spectral flow transformation. The Ramond-Ramond (RR) supersymmetric sector of the theory contains the states that are relevant for the statistical ensemble of states describing the black hole in type IIB supergravity on $S^1\times\mathcal{M}_4$ \cite{David:2002wn, Mathur:2005zp, Giusto:2013bda, Giusto:2013rxa, Bena:2015bea, Giusto:2015dfa, Bena:2016ypk, Bena:2017xbt, Bakhshaei:2018vux}.

In this setup, the four-point function we focused on, often dubbed as Heavy-Heavy-Light-Light (HHLL) correlator, is of the form
\begin{equation}
\label{eq:1}
    \langle O_H(z_1,\bar{z}_1)\bar{O}_H(z_2,\bar{z}_2)O_L(z_3,\bar{z}_3)\bar{O}_L(z_4,\bar{z}_4)\rangle \,,
\end{equation}
where the heavy operators ($O_H$) have conformal dimensions scaling with the central charge $c$, while the light ones ($O_L$) have dimensions of order unit. From the gravity point of view the heavy states we will focus on are described by smooth, horizonless solutions of type IIB on ${\cal M}_4$, while the light probes are dual to some perturbations around this heavy background. When the D1D5 CFT is at the free point it is possible to calculate the correlator \eqref{eq:1} with standard technique, while in the opposite limit we have to use holographic methods. 

The class of correlators introduced above has recently been studied in \cite{Galliani:2016cai, Galliani:2017jlg, Bombini:2017sge} for two-charge microstates, corresponding to $\frac{1}{4}$-BPS RR ground states, whose statistical ensemble is not dual to a macroscopic black hole at the level of classical gravity, but it provides a good testing ground as we know in detail all the gravitational solutions dual to these states \cite{Kanitscheider:2006zf, Kanitscheider:2007wq}. We want now to take a step further and study HHLL correlators for a large class of three-charge microstates recently found \cite{Bena:2016ypk, Bena:2017xbt}, and whose thermodynamic description is actually a black hole with a macroscopic entropy already at the level of classical gravity. These microstates break half of the supersymmetries of the two-charge seed solution and are therefore $\frac{1}{8}$-BPS states and they are schematically written in the RR sector as
\begin{equation}
    \label{eq:2}
    \left(\left|++\right\rangle_1\right)^{N_1}\prod_{k,m,n}\left(\frac{\left(J^+_{-1}\right)^m}{m!}\frac{\left(L_{-1}-J_{-1}^3\right)^n}{n!} |00\rangle_k\right)^{N_{k,m,n}} \, ,
\end{equation}
where $L_{n}$, $J^a_{n}$ are the generators of the Kac-Moody algebra of the CFT. We analyze also another class of three-charge microstates very recently found in \cite{Ceplak:2018pws} (for other three-charge geometries, see \cite{Mathur:2011gz, Giusto:2012yz, Lunin:2012gp, Bena:2017upb, Bena:2018mpb, Heidmann:2019zws}). They are constructed by acting also with the supercharges on the two-charge seed, spectrally-flowed into the NS sector, and they read 
\begin{equation}
\left|0\right\rangle_1^{N_1}\prod_{k,m,n,q}\left(\frac{\left(J^+_{0}\right)^m}{m!}\frac{\left(L_{-1}\right)^n}{n!} \left(   G_{-\half}^{+1}  G_{-\half}^{+2} + \frac{1}{k} \, J_0^3 L_{-1}  \right)^q |O^{--}\rangle_k\right)^{N_{k,m,n,q}} \,.
\end{equation}
The numbers $N_1$, $N_{k,m,n}$ (or $N_{k,m,n,q}$) controls the number of each \textit{strand} in the D1D5 orbifold picture, and in gravity corresponds to some parameters $a$ and $b$ whose strength controls the depth of the throat of the microstates. The infinite throat limit corresponds to take the limit $a^2/b^2\to 0$. Even though exact results in the parameters $a$ and $b$ have been found in \cite{Bombini:2017sge} for a class of two-charge geometries, for the three-charge state considered here, we restrict to the regime where we take the ratio $b^2/a^2$ to be small. 

Notice that it is not straightforward to use Witten diagrams to calculate the correlators \eqref{eq:1} since the heavy states correspond to multi-particle operators with a large conformal dimension and are not dual to a single supergravity mode. We bypass these difficulties by seeing the four-point function as a two-point function in a non-trivial background state. In the gravity picture, this boils down to solve a wave equation obtained by perturbing the fields dual to the light operators, around the known smooth geometries, dual to heavy states. The results obtained pass a set of non-trivial consistency checks and they can be related to each other, as well as to the two-charge results obtained in \cite{Galliani:2017jlg, Bombini:2017sge}, by a set of Ward Identities (WI)  encoding, in form of differential operators, the action of all the generators of the global part of the superconformal algebra on the two-charge states. 

Moreover, a very recent conjecture put forward in \cite{Giusto:2018ovt}, has been employed to extract the all-light (LLLL) correlators from the corresponding HHLL version. In fact, a class of LLLL four-point functions has been constructed in \cite{Giusto:2018ovt} starting from the HHLL correlator computed in the two-charge geometries. In analogy to that work, we will also go towards the extraction of the LLLL version of our HHLL four-point function. The LLLL correlators can be thus interpreted in terms of a sum of Witten diagrams, each of them reflects the exchange of fields in AdS, in different channels. The prescription of \cite{Giusto:2018ovt} allows to extract the $s$-channel of the LLLL correlator straightforwardly from the HHLL one. As it will be more clear in the following, in some cases it will be possible to reconstruct all the other channels in order to get the entire LLLL four-point function. Fundamental tools in this approach are the Ward Identities, relating our three-charge correlators to the two-charge ones whose LLLL version is known. Furthermore, as recently developed in AdS/CFT context, the Mellin formalism \cite{Penedones:2010ue, Rastelli:2016nze, Rastelli:2017udc} constitutes a natural language in which holographic correlators can be interpreted, and it will turn out to be fundamental to analyze the dynamical properties of our results, besides providing another non-trivial consistency check of our results.

The paper is organized as follows: in Section \ref{sec:CFT}, we discuss the CFT picture, providing detailed definitions of the essential ingredients we need; Section \ref{sec:gravity} contains the dual gravity picture for every case we will focus on, explicitly writing down the geometry dual to the state considered in the CFT. The same section contains also the holographic computations and results for the for-point functions. In Section \ref{sec:LLLL} we will use the method of \cite{Giusto:2018ovt} to extract the $s$-channel of the LLLL version of the HHLL four-point function computed in the sec.~\ref{sec:gravity}. In Section \ref{sec:WI} we will find the Ward identities that we will use to connect our results to the correlators containing two-charge states; those will be also used to find the other channel in order to reconstruct the entire LLLL three-charge four-point functions. We will end the paper with a discussion in Section \ref{sec:disc}. The appendices contain technicalities that are helpful to find our results.

\section{CFT picture}
\label{sec:CFT}
In this section we use the D1D5 CFT at the free orbifold point to describe the correlators under analysis, defining the heavy and the light operators we will use, following the notation of \cite{Avery:2010qw, Giusto:2015dfa}, briefly reviewed in app.~\ref{sec:AppA}. At the orbifold point, the CFT target space is $(\mathcal{M}_4)^N/S_N$ (where $\mathcal{M}_4$ can be either $T^4$ or $K_3$) and the theory can be formulated in terms of $N$ groups of free bosonic and fermionic fields
\begin{equation}
  \label{eq:ff}
  \Big(\partial X^{A\dot{A}}_{(r)}(z),\, \psi^{\alpha \dot{A}}_{(r)} (z)\Big),~~~
  \Big(\bar\partial X^{A\dot{A}}_{(r)}(\bar z),\, \tilde\psi^{\dot{\alpha} \dot{A}}_{(r)} (\bar z)\Big),
\end{equation}
where $(A, \dot{A})$ is a pair of $SU(2)$ indices forming a vector in the CFT target space, while $(\alpha,\dot{\alpha})$ are indices of $SU(2)_L\times SU(2)_R$ which is part of the R-symmetry group; finally $r=1,\ldots N$ is a flavour index running on the various copies of the target space on which the symmetric group $S_N$ acts. 
The algebra of the theory in all points of the moduli space is given by an affine algebra generated by three conserved currents
\begin{equation}
  \label{eq:alg}
  \Big(T(z),\, J^a(z),\, G^{\alpha A}(z)\Big),\quad \Big(\bar{T}(\bar{z}),\, \bar{J}^a(\bar{z}),\, \bar{G}^{\alpha A}(\bar{z})\Big) ,
\end{equation}
where $T(z)$ is the stress energy tensor generating the conformal transformations, $J^a(z)$ is the $SU(2)_L$ R-symmetry current and $G^{\alpha A}(z)$ is the supercurrenct. 

As mentioned in the introduction, we will study four-point functions with two light operators and two heavy operators. In order to identify and define these four operators we remind that the spectrum of the theory usually decomposes in two sectors, given by the two different periodicity of the fermionic fields under rotation: the Neveu-Schwarz (NS) and the Ramond (RR) sector. On the CFT side the operation of going to one of these sectors to the other one is implemented via a spectral flow transformation that has a particular action on the operators of the theory (see sec.~\ref{sec:WI} for details). 
Moreover, for each of these sectors the spectrum further decomposes as a sum over twisted sectors, that are in one-to-one correspondence to the conjugacy class in the symmetric group $S_N$. When $k$ copies of the orbifold CFT are sewn together by a cyclic permutation of boundary condition, the operators have to be diagonalized in the sense explained in details in app.~A of \cite{Bombini:2017sge}.

The object we will focus on is then 
\begin{equation}
\label{eq:4pf}
    \langle O_H(z_1,\bar{z}_1)\bar{O}_H(z_2,\bar{z}_2)O_L(z_3,\bar{z}_4)\bar{O}_L(z_4,\bar{z}_4)\rangle=\frac{1}{z_{12}^{2h_H}z_{34}^{2h_L}}\frac{1}{\bar{z}_{12}^{2\bar{h}_H}\bar{z}_{34}^{2\bar{h}_L}} \, \mathcal{G}(z,\bar{z})\,,
\end{equation}
where $\mathcal{G}$ is a function of the cross ratios
\begin{equation}
    \label{eq:cross}
    z=\frac{z_{14}z_{23}}{z_{13}z_{24}},\quad \bar{z}=\frac{\bar{z}_{14}\bar{z}_{23}}{\bar{z}_{13}\bar{z}_{24}} \,,
\end{equation}
and $z_{ij}=z_i-z_j$. In order to easily find the function $\mathcal{G}$ one can perform a conformal transformation fixing three of the four points, let us say $z_2\to\infty,\,z_1=0$ and $z_3=1$, which further implies $z=z_4$:
\begin{equation}
    \label{eq:4pf2}
    \langle\bar{O}_H|O_L(1)\bar{O}_L(z,\bar{z})|O_H\rangle\equiv\mathcal{C}(z,\bar{z})=\frac{1}{(1-z)^{2h_L}}\frac{1}{(1-\bar{z})^{2\bar{h}_L}} \, \mathcal{G}(z,\bar{z}) .
\end{equation}

In what follows, we are going to define which operators we will consider in computing the correlators defined above. 
We will concentrate on supersymmetric ground states as heavy states, whose dual gravity solutions are known \cite{Giusto:2013bda, Bena:2015bea, Giusto:2015dfa, Bena:2016ypk, Bena:2017xbt, Bakhshaei:2018vux} and, as light operators, we will consider a class of operators dual to a family of perturbation of the fields that are minimally coupled massless scalars in gravity side.

A class of important heavy states are known to be $\frac{1}{4}$-BPS states consisting in short multiplets labelled by spin doublet $\alpha,\,\dot{\alpha}$ under R-symmetry, and $A,\,B$ under auxiliary $SU(2)$:
\begin{equation}
  \label{eq:2ch}
  |\alpha\dot{\alpha}\rangle_k,\,|AB\rangle_k,\,|\alpha B\rangle_k,\,|A\dot{\alpha}\rangle_k \, .
\end{equation}
Since they are going to play a central role in the following, it is worth to define the states $\left|++\right\rangle_k$ as the highest-weight state of R-symmetry with eigenvalues $(\jmath, \bar \jmath) = (+\half, + \half)$ under $(J^3, \tilde J^3)$, and the state $|00\rangle_k$ as a particular combination of the states described above, defined as $|00\rangle_k\equiv\epsilon^{AB}|AB\rangle_k$, that has instead $(\jmath, \bar \jmath) = (0,0)$.

The full state in the orbifold theory is then a tensor product of ground states for the cyclic twists in the
symmetric group conjugacy class, having $N(s)$ copies of $k$-cycle ground states \eqref{eq:2ch} of the polarization state s. The class of state then takes the form
\begin{equation}
  \label{eq:2chfull}
 \psi_{\{N_k^s\}}\equiv \prod_{k,s}\left(|s\rangle_k\right)^{N_k^s}  \, .
\end{equation}
These are usually called two-charge states and their gravity dual are known and completely classified \cite{Kanitscheider:2006zf, Kanitscheider:2007wq}.

Another class of heavy states, which are the ones we will focus on, are three-charges, $\frac{1}{8}$-BPS state and are given by 
\begin{equation}
  \label{eq:3ch}
 \psi_{\{N_1,N_{k,m,n}\}}\equiv \left(\left|++\right\rangle_1\right)^{N_1}\prod_{k,m,n}\left(\frac{\left(J^+_{-1}\right)^m}{m!}\frac{\left(L_{-1}-J_{-1}^3\right)^n}{n!} |00\rangle_k\right)^{N_{k,m,n}} \, .
\end{equation}
The integer numbers $\{N_1,N_{k,m,n}\}$ specify the number of strands with particular quantum numbers and must satisfy
\begin{equation}
  \label{eq:cond}
 N_1+\sum_{k,m,n}kN_{k,m,n}=N \,.
\end{equation}

In order to have a dual classical supergravity solution for these states we have to take a coherent superposition of them. In particular the heavy states we are interested in, with a gravity dual are given by
\begin{equation}
\label{eq:3chfull}
  |k,m,n\rangle\equiv\sum_{N_1,N_{k,m,n}}A_1^{N_1}\left(B_{k,m,n}\right)^{N_{k,m,n}}\psi_{\{N_1,N_{k,m,n}\}} \,,
\end{equation}
where the sum is restricted to $\{N_1,N_{k,m,n}\}$ satisfying \eqref{eq:cond}, that gives the condition
\begin{equation}
    \label{eq:cond2}
    |A_1|^2+\binom{k}{m}\binom{n+k-1}{n}|B_{k,m,n}|^2=N \, .
\end{equation}
It has been proposed in \cite{Bena:2016ypk, Bena:2017xbt} that the states \eqref{eq:3chfull} are the holographic dual of a class of single-mode supergravity solution whose explicit form can still be found in the same work. We will describe these dual solutions in sec.~\ref{sec:gravity}.

In particular, we will focus on two classes of heavy states
\begin{equation}
O_H\to \,|O_H\rangle\equiv \left\{|1,0,n\rangle,\, |m,m,0\rangle\right\} \,,
\end{equation}
as defined in \eqref{eq:3chfull} with the condition
\begin{equation}
   |A|^2+|B|^2=N \,, 
\end{equation}
where we have defined $A\equiv A_1$ and $B\equiv\left\{B_{1,0,n},\,B_{m,m,0}\right\}$.

In \cite{Ceplak:2018pws} the family discussed above was enlarged. In the NS sector, where, as we will discuss in sec.~\ref{sec:WI}, $(L_{-1} - J_{-1}^3)  \mapsto L_{-1}$ and $J_{-1}^+ \mapsto J_0^+$, we can act with another linearly independent operator constructed with the supercurrents, to build the enlarged family 
{\small
\begin{equation}
    \psi_{\{N_1,N_{k,m,n,q}\}}^{\rm NS} \equiv \left|0\right\rangle_1^{N_1}\prod_{k,m,n,q}\left(\frac{\left(J^+_{0}\right)^m}{m!}\frac{\left(L_{-1}\right)^n}{n!} \left(   G_{-\half}^{+1}  G_{-\half}^{+2} + \frac{1}{k} \, J_0^3 L_{-1}  \right)^q |O^{--}\rangle_k\right)^{N_{k,m,n,q}} \, ,
\end{equation} 
}%
where $q=0,1$ and where again the integer numbers $\{N_1,N_{k,m,n,q}\}$ specify the number of strands with particular quantum numbers and must satisfy
\begin{equation}
  \label{eq:condq}
 N_1+\sum_{k,m,n,q } k N_{k,m,n,q}= N \,.
\end{equation}
In order to have a dual classical supergravity solution for these states, we have again to take a coherent superposition of them. In particular the heavy states we are interested in are given by
\begin{equation}
\label{eq:3chfullq}
  |k,m,n, q\rangle\equiv\sum_{N_1,N_{k,m,n,q}}A_1^{N_1}\left(B_{k,m,n,q}\right)^{N_{k,m,n}}\psi_{\{N_1,N_{k,m,n, q}\}} \,,
\end{equation}
where the sum is restricted to $\{N_1,N_{k,m,n,q}\}$ satisfying \eqref{eq:condq}. To ease the notation, when referring to states with $q=0$, we will always drop it from the nomenclature, i.e. we will refer to $|1,0, n \ket$ or $|m,m,0\ket$ states, not to $|1,0, n ,0\ket$ nor $|m,m,0,0\ket$ states. With this family, we have listed a wide family of the $\frac{1}{8}$-BPS states whose dual geometries are explicitly known as type IIB supergravity solutions on AdS$_3\times S^3 \times {\cal M}_4$.

For what concerns the light operators we will work with the following ones
\begin{equation}
    \label{eq:light}
    O_L\to O_{\text{bos}}=\sum_{r=1}^N\frac{\epsilon_{\dot{A}\dot{B}}}{\sqrt{2N}}\partial X^{1\dot{A}}_{(r)}\bar{\partial}X^{1\dot{B}}_{(r)},\quad \bar{O}_L\to \bar{O}_{\text{bos}}=\sum_{r=1}^N\frac{\epsilon_{\dot{A}\dot{B}}}{\sqrt{2N}}\partial X^{2\dot{A}}_{(r)}\bar{\partial}X^{2\dot{B}}_{(r)}.
\end{equation}

With this choice of light and heavy operators (in the case $q=0$), the correlator at the orbifold point depends only on the strand structure, but not on the particular quantum numbers of the RR ground state considered. A standard way to calculate this correlator is to diagonalize the boundary conditions and then to take the linear combination of the contributions of each strand as done in \cite{Bombini:2017sge}. We report the results for first two cases, where the heavy states are in the untwisted sector and the strand structure is trivial
\eqref{eq:light}
\begin{equation}
    \label{eq:orb1}
    \mathcal{C}_{(1,0,n)}=\mathcal{C}_{(m,m,0)}=\frac{1}{|1-z|^4}  \, .
\end{equation}
The computation for $\mathcal{C}_{(2,0,0,1)}$ instead is more involved and, since it is not relevant for the aim of the present paper, we avoid to report it here. In all cases we study, we will see that at the strong coupling point the correlators differ from the ones computed at the free point.

\section{Gravity picture and Holographic Correlators in $\frac{1}{8}$-BPS state}
\label{sec:gravity}


We now introduce the $k,m,n$ geometries built in \cite{Bena:2016ypk, Bena:2017xbt, Bakhshaei:2018vux}; those are type IIB supergravity solutions described by four scalar functions $Z_1$, $Z_2$, $Z_4$ and ${\cal F}$, three two-forms $\Theta_1$, $\Theta_2$ and $\Theta_4$, and a 1-form  $\omega$, by which we can describe all the type IIB fields:
\begin{subequations}\label{eq:FullMetrickmn}
\begin{align}
    ds^2 &= \sqrt{\frac{Z_1 Z_2}{\cal P}} \, ds_6^2 + \sqrt{\frac{Z_1}{Z_2}}\, ds_{T^4} \,, \\
    ds_6^2 &= - \frac{2}{\sqrt{\mathcal{P}}} (dv+\beta) \left[ du+\omega + \half \F (dv+\beta) \right] + \sqrt{\mathcal{P}} \, ds_4^2 \, , \\
    ds_4^2 &= \Sigma \left( \frac{dr^2}{r^2+a^2} + d\theta^2\right) + (r^2+a^2) \sin^2 \theta \, d \phi^2 + r^2 \cos^2 \theta \, d\psi^2 \,, \\
\Sigma &= r^2 +a^2 \cos^2 \theta\,, \quad {\cal P} = Z_1 Z_2 - Z_4^2 \,, \quad u = \frac{t-y}{\sqrt{2} } \,, \quad v = \frac{t+y}{\sqrt{2} }  \\
\beta &=  \frac{R\, a^2}{\sqrt{2}\,\Sigma} \left(\sin^2 \theta \, d\phi - \cos^2 \theta \, d\psi\right)  \,,
\end{align}
\end{subequations}
where we are focusing only on the metric, since it is the only field that will be relevant for us. The objects defining this ansatz have to satisfy two layers of differential equations, in order to have a solution of the type IIB equations of motion:
\be\label{eq:layer1}
\begin{split}
*_4 \D \dot Z_1 & = \D \Theta_2 \,, \quad \D *_4 \D Z_1 = - \Theta_2 \wedge d\beta \,, \quad \Theta_2 = *_4 \Theta_2 \,, \\
*_4 \D \dot Z_2 & = \D \Theta_1 \,, \quad \D *_4 \D Z_2 = - \Theta_1 \wedge d\beta \,, \quad \Theta_1 = *_4 \Theta_1 \,, \\
*_4 \D \dot Z_4 & = \D \Theta_4 \,, \quad \D *_4 \D Z_4 = - \Theta_4 \wedge d\beta \,, \quad \Theta_4 = *_4 \Theta_4 \,, \\
\end{split}
\ee
and 
\be\label{eq:layer2}
\begin{split}
\D \omega + *_4 \D \omega_4 + {\cal F} \, d\beta &= Z_1 \Theta_1 + Z_2 \Theta_2 - 2 Z_4 \Theta_4 \,, \\
*_4 \D *_4 \left( \dot \omega - \half \D {\cal F} \right) &= \pd_v^2 (Z_1 Z_2 - Z_4^2) - [\dot Z_1 \dot Z_2 - (\dot Z_4)^2] - \half *_4 (\Theta_1 \wedge \Theta_2 - \Theta_4 \wedge \Theta_4 ) ,
\end{split}
\ee
where we have defined $\D = d_4 - \beta \wedge \pd_v$. The $k$, $m$, $n$ family of solutions built in \cite{Bena:2016ypk, Bena:2017xbt} is described by 
\begin{equation}
    \begin{split}
        Z_1 &= \frac{Q_1}{\Sigma} + \frac{R^2}{Q_5} \, \frac{b^2}{2} \, \frac{\Delta_{2k,2m,2n}}{\Sigma} \, \cos \hat v_{2k, 2m, 2n} \,, \quad Z_2 = \frac{Q_5}{\Sigma} \,, \quad Z_4 = R \,  b \, \frac{\Delta_{k,m,n}}{\Sigma} \, \cos \hat v_{k,m,n} \,,\\
        \Theta_1 &=0 \,, \quad \Theta_2 = \frac{R}{Q_5} \, \frac{b^2}{2} \, \vartheta_{2k,2m,2n} \,, \quad Z_4 = b \,  \vartheta_{k,m,n} \,,
    \end{split}
\end{equation}
where we have defined 
\begin{subequations}
\begin{align}
    \Delta_{k,m,n} &= \left( \frac{a}{\sqrt{r^2+a^2}} \right)^k \left( \frac{r}{\sqrt{r^2+a^2}} \right)^n \cos^m \theta \, \sin^{k-m} \theta \,, \\
    \hat v_{k,m,n} &= (m+n) \frac{\sqrt{2}\, v}{R} + (k-m) \phi - m \psi \,, 
\end{align}
\end{subequations}
while $\vartheta_{k,m,n}$ is defined in eq. (3.20) of \cite{Bena:2017xbt}. We do not report here its precise form, since it will not be useful nor relevant. In order to have non-singular geometries we need to impose a regularity condition 
\begin{equation}
    a^2 + x_{k,m,n} \, \frac{b^2}{2}  = \frac{Q_1 Q_5}{R^2} \equiv a_0^2 \,, \quad x^{-1}_{k,m,n} = {k \choose m} {k+n-1 \choose n} .
\end{equation}

The relation between $a$, $b$ in the gravity side with the $A$, $B$ of eq. \eqref{eq:cond2} in the CFT side is \cite{Giusto:2015dfa}
\begin{equation}
    |A|= R \sqrt{\frac{N}{Q_1 Q_5}} \, a \,, \quad |B| = R \sqrt{\frac{N}{2 Q_1 Q_5}}\, x_{k,m,n} \, b \,.
\end{equation}

The missing ${\cal F}$, $\omega$ can be computed via the second layer of equations.  A close form for them is known only for particular choices of the three parameters \cite{Bena:2016ypk, Bena:2017xbt, Bakhshaei:2018vux, Ceplak:2018pws}; we will use here only two of those choices: the $(k,m,n) = (1,0,n)$ and the $(k,m,n)=(m,m,0)$, that we discuss in detail in the following.

It can be proven that, in all the three-charge geometries \eqref{eq:FullMetrickmn}, the supergravity field dual to our operator \eqref{eq:light} is a minimally coupled massless scalar field in six dimensions with $Y^{00}$ harmonic\footnote{For the harmonic functions on $S^3$ we use the notation of \cite{Galliani:2017jlg}.} on the $S^3$ (see app.~B of \cite{Bombini:2017sge}), i.e. 
\begin{equation}
    \square_6 \left( B(\tau, \sigma, r) Y^{00} (\theta, \phi, \psi) \right) = 0 \,,
\end{equation}
where $\square_6$ is the scalar laplacian of the $ds_6^2$ metric, i.e. $\square_6 \cdot = \frac{1}{\sqrt{g_6}} \pd_M (\sqrt{g_6} g_6^{MN}\pd_N \cdot )$, and where $\tau = t/R$, $\sigma = y/R$. We then resort to standard holographic methods to extract the Heavy-Light four-point function; In fact, we can compute it by solving the equation of motion for the dual supergravity field with the appropriate boundary conditions, i.e. 
\begin{equation}
    B(\tau, \sigma, r) \sim \delta(\tau , \sigma) + \frac{b(\tau, \sigma)}{r^2} \,,
\end{equation}
for $r\to \infty$, plus regularity at $r=0$. From here we will read the correlator as
\begin{equation}
    \bra O_H (0) \bar{O}_H (\infty) O_L (1) \bar{O}_L (z, \bar z) \ket = |z|^{-2} \, b(z, \bar z) \,,
\end{equation}
where we have mapped the cylinder to the plane via $z=e^{\tau_E + i \sigma}$, $\bar z = e^{\tau_E - i \sigma}$, with $\tau_E = i \tau$.

More precisely, we will resort to a perturbative solution of the equation, expanding in powers of $\frac{b^2}{2a_0^2}$, that gives a separable equation of motion, since all the $(k,m,n)$ geometries approaches the vacuum AdS$_3 \times S^3$ solution when $b\to 0$. It will be then useful to rewrite the metric in the following form
\begin{equation}
    ds_6^2 = V^{-2} g_{\mu\nu}dx^\mu dx^\nu + G_{ab} (d\theta^a + A^a) (d\theta^b + A^b) , \quad V^2 = \frac{\det G_{ab}}{ \sin^2 \theta \cos^2 \theta} \,,
\end{equation}
where $A^a = A^a{}_\mu dx^\mu$ can be seen as a 1-form on AdS$_3$. We have split the six-dimensional coordinates as $x^M = (x^\mu , \theta^a)$ where $x^\mu =\{ \tau, \sigma, r\}$ and $\theta^a = \{ \theta, \phi, \psi\}$. Schematically, we will have then
\begin{equation}
    \begin{split}
        \square_6 \left[\left( B_0 +   \frac{b^2}{2a_0^2} \, B_1 \right) Y^{00}\right] \simeq \square_0 B_0 + \frac{b^2}{2a_0^2} \left( \square_0 B_1  + \frac{b^2}{2a_0^2} \, \square_1 B_0 \right) + {\cal O}\left( \frac{b^4}{4a_0^4} \right),
    \end{split}
\end{equation}
where $\square_0 \cdot = \frac{1}{\sqrt{g_3}} \pd_M (\sqrt{g_3} g_3^{MN}\pd_N \cdot )$ is the scalar laplacian of AdS$_3$ in global coordinates; this equation can be formerly solved order by order; at the zeroth order, the solution that respects the correct boundary solutions is the AdS$_3$ Bulk-to-Boundary propagator:
\begin{equation}
    B_0 (\tau, \sigma, r) = K_2^{\rm glob} ( \tau, \sigma, r |\tau'=0, \sigma'=0 ) = \left[ \half \, \frac{a_0}{\sqrt{r^2+a_0^2} \, \cos \tau - r \cos \sigma }  \right]^2 .
\end{equation}
The second order can be computed via the Green-function method
\begin{equation}
 B_1 (\tau, \sigma, r) = - i \int d^3 r' \sqrt{-\bar{g}_3} \, G({\bf r} | {\bf r}')  J_s ({\bf r}')  \,, \quad J_s \equiv - \bra  \square_1 B_0 \ket \,,
\end{equation}
since at the $b^0$ order the metric reduces to AdS$_3 \times S^3$. Now, from the large $r$-limit of $B_1$, that is deduced by the large $r$-limit of the Bulk-to-Bulk propagator\footnote{Here we explicitly use that $\Delta_L = 2$.} as $G_2 ({\bf r}' | r,\tau, \sigma) \to \frac{a_0^2}{2\pi r^2} \, K_2({\bf r}' |\tau, \sigma)$, we get, mapping onto the plane,
\begin{equation}
    \left. \bra O_H (0) \bar{O}_H (\infty) O_L (1) \bar{O}_L (z, \bar z) \ket \right|_{b^2} =  - \frac{b^2}{2\pi}  \int d^3 w' \sqrt{-\bar{g}_3} \, K_2 (  {\bf w}' | z , \bar{z} )  J_s ({\bf w}') \,.
\end{equation}

\begin{figure} 
    \centering
    \begin{tikzpicture}[scale=1]
    \draw [thick, blue, double](1,1) node[above]{$\phantom{1} \hspace{2mm} \textcolor{blue}{O_H}$} --(0,0) -- (1,-1) node[below]{$\phantom{1} \hspace{2mm}  \textcolor{blue}{\bar{O}_H} $};
    \draw (-1,-1) node[below]{$  \bar{O}_L \hspace{2mm} \phantom{1}$}  --(0,0) -- (-1,+1) node[above]{$O_L \hspace{2mm} \phantom{1}$};
    \draw [thick] (0,0) circle [radius=1.414];
    \draw [fill, white] (0,0) circle [radius=0.55]; 
    \draw [pattern=horizontal lines, thick] (0,0) circle [radius=0.55]; 
    \end{tikzpicture}
    \centering
    \begin{tikzpicture}[scale=1] 
    \draw [draw=white, thin] (0,-0.01)-- (0,0) node[baseline]{$=$}  -- (0,+0.01);
    \draw [draw=white, thin] (0,1.7) -- (0,-1.7);
    \draw [draw=white, thin] (0.4, 0) --(-0.4,0);
    \end{tikzpicture}
    \centering
    \begin{tikzpicture}[scale=1]
    \draw [thick] (0,0) circle [radius=1.414];
    \draw  (-1,+1) node[above]{$O_L \hspace{2mm} \phantom{1}$} -- (-0.4, 0) -- (-1,-1) node[below]{$\bar{O}_L \hspace{2mm} \phantom{1}$};
    \draw [photon, thick] (-0.4, 0) -- (+0.4, 0);
    \draw [ultra thick, blue] (0.6, 0) circle [radius=0.2];
    \draw [blue, thick] (0.4586, -0.1414) -- (0.6, 0) node[below=6]{$ \phantom{1} \vspace{20mm} O_H$} -- (0.7414, 0.1414);
    \draw [blue, thick] (0.7414, -0.1414) -- (0.6, 0) -- (0.4586, 0.1414);
    \end{tikzpicture}
    \caption{A pictorial representation of the method to compute the HHLL four-point function, seen as a two-point function of the light operators on a  non-trivial background sourced by the heavy operators. In the left-hand side, the dual supergravity fields of the single-trace light operators are represented by black straight lines in the bulk; on the contrary the heavy operators, being multi-trace, do not have a representation in terms of single-mode supergravity fields, and then their duals in the bulk are therefore pictorially represented by a blue double-line. In the right-hand side instead, we represent the fact that heavy states source a non-trivial geometry acting like a background field, represented by a crossed blue circle.   }
    \label{fig:HHLLWD}
\end{figure}
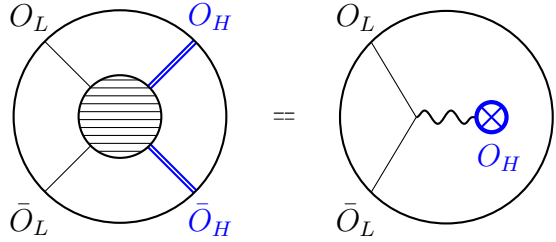

This method for computing the four-point function allows us to avoid to use the Witten Diagram technology, that is still not properly defined for AdS$_3$ \cite{Giusto:2018ovt}, since only the cubic coupling have been worked out \cite{Arutyunov:2000by}. To have an intuitive picture in mind, we report in fig.~\ref{fig:HHLLWD} a graphic representation of how this method allows us to compute the Heavy-Light four-point function.

\subsection{HHLL 4-point functions in the $(k,m,n)=(1,0,n)$ geometry}
\label{subsec:10n}


The explicit solution of the two layers (\ref{eq:layer1}, \ref{eq:layer2}) for the $(k,m,n)=(1,0,n)$ geometry was found in \cite{Bena:2016ypk, Bena:2017xbt}
\begin{subequations}
\begin{align}
ds_6^2 &= - \frac{2}{\sqrt{\mathcal{P}}} (dv+\beta) \left[ du+\omega + \half \F (dv+\beta) \right] + \sqrt{\mathcal{P}} \, ds_4^2 \, ,\\
\hat v_{1,0,n} &= \frac{\sqrt{2}}{R} \, n v + \phi\,,\quad  \hat v_{2,0,2n} = \frac{\sqrt{2}}{R} \,2 n v +2 \phi\, ,  \\
\Delta_{1,0,n} &= \frac{a \, r^n}{(r^2+a^2)^{\frac{n+1}{2}}}  \, \sin \theta \, , \quad \Delta_{2,0,2n} =  \frac{a^2 \, r^{2n}}{(r^2+a^2)^{n+1}}  \, \sin^2 \theta \\
Z_1 &= \frac{Q_1}{\Sigma} + \frac{R^2}{2 Q_5} \, b^2 \, \frac{\Delta_{2,0,2n}}{\Sigma} \cos \hat v_{2,0,2n} \, \quad Z_2 = \frac{Q_2}{\Sigma}\, , \quad Z_4 = b R \, \frac{\Delta_{1,0,n}}{\Sigma}\cos \hat v_{1,0,n} \\
\omega &= \frac{a^2 R}{\sqrt{2}\,\Sigma} \left( \sin^2 \theta \, d\phi + \cos^2 \theta \, d\psi \right)+ \frac{b^2}{a^2}\frac{a^2 R}{\sqrt{2}\,\Sigma} \left[ 1- \frac{r^{2n}}{(r^2+a^2)^n}   \right] \sin^2 \theta \, d\phi \,, \\
 \F &= -\frac{b^2}{a^2} \left[ 1- \frac{r^{2n}}{(r^2+a^2)^n} \right] \, , \quad\beta = \frac{a^2 R}{\sqrt{2}\,\Sigma} \left( \sin^2 \theta \, d\phi - \cos^2 \theta \, d\psi \right) \, , 
\end{align}
\end{subequations}
so that, calling $F_n \equiv  \left[ 1- \frac{r^{2n}}{(r^2+a^2)^n} \right]$, 
\be
\omega =  \frac{a^2 R}{\sqrt{2}\,\Sigma} \left[ \left( 1+ \frac{b^2}{a^2} F_n  \right) \sin^2 \theta \, d\phi + \cos^2 \theta \, d\psi \right]  \, .
\ee

Notice that this is a 3-charge geometry, due to the fact that we have a non-vanishing ${\cal F}$, controlled by a non-vanishing $n$; sending $n\to 0$ will reduce it to a 2-charge geometry, as expected. This metric can be recast, via the splitting $x^M = (x^\mu, \theta^a)$, with $M=0,\ldots, 5$, $\mu=0,1,2$ and $a=3,4,5$, as
\be
ds_6^2 = V^{-2} g_{\mu\nu} dx^\mu dx^\nu + G_{ab}(d\theta^a + A^a_\mu dx^\mu) (d\theta^b + A^b_\nu dx^\nu) , 
\ee
where $V$ is fixed requiring that $\sqrt{-\det G_{MN}} = \sqrt{-\det g_{\mu\nu}} \sqrt{\det q_{ab}}$ with $q_{ab}$ the round $S^3$ metric, for all values of $b$. The three-dimensional non-compact metric is
\be\label{3charge_metric_6}
\begin{split}
ds_3^2 &= - \left[r^2 \left(1 - \frac{b^2}{2a_0^2} \, F_n \right) + \frac{a^4}{a_0^2} \right] d\tau + r^2  \left(1 + \frac{b^2}{2a_0^2} \, F_n \right) d \sigma^2 \\
& \quad \  +\frac{r^2 + \frac{a^2}{a_0^2}\left( a^2 + \frac{b^2}{2} \, F_n \right)}{(r^2+a_0^2)^2} dr^2 \,,
\end{split}
\ee
where we recall that we have defined
\be
\tau = \frac{t}{R} \,, \quad \sigma = \frac{y}{R}\,.
\ee
We then try to solve perturbatively the scalar equation, as explained previously, giving us the source 
\be
\begin{split}
J_s = \, &  \frac{R^2 }{2(r^2+a_0^2)^2}\left[(r^2+2a_0^2) - \frac{r^{n}}{(r^2+a_0^2)^{n-1}}  \right] \pd_t^2 B_0  - \frac{  R^2 }{r^2+a^2_0}\left[ 1- \frac{r^{n}}{(r^2+a_0^2)^{n}} \right] \pd_t \pd_y B_0 \\
& + \half \left( \frac{1}{r^2} -  \frac{r^{n}}{(r^2+a_0^2)^{n-1}}  \right) R^2 \pd_y^2 B_0 + \frac{1}{r}\, \pd_r  \left( r \pd_r B_0 \right) \, .
\end{split}
\ee
In the case $n=0$ we recover the result of the 2-charge case
\be
J_s^{(n=0)} = \frac{a_0^2 R^2}{2(r^2+a_0^2)^2} \,  \pd_t^2 B_0  +  \frac{a_0^2 R^2}{2r^2(r^2+a_0^2)} \,  \pd_y^2 B_0  + \frac{1}{r}\, \pd_r  \left( r \pd_r B_0 \right) \, .
\ee
After some simple algebraic manipulations, we can rewrite the source as\footnote{Notice that, since
$$
B_{\pm} = \frac{a_0}{\sqrt{r^2+a^2_0}} \, e^{\pm i \tau} \,,
$$
we have
$$
\frac{R^2}{r^2+a_0^2} =\frac{R^2}{a_0^2} ( B_-^{(1)}B_+^{(1)}) \, .
$$}  
\begin{subequations}
\begin{align}
    J_s &= J_s^{(n=0)} + J_s^{(n>0)} \,, \\
    J_s^{(n=0)} &= \frac{a_0^2 R^2}{2(r^2+a_0^2)^2} \,  \pd_t^2 B_0  +  \frac{a_0^2 R^2}{2r^2(r^2+a_0^2)} \,  \pd_y^2 B_0  + \frac{1}{r}\, \pd_r  \left( r \pd_r B_0 \right) \, , \\
    J_s^{(n>0)} &=  \frac{R^2}{a_0^2} \, B_+^{(0)} B_-^{(0)} \left[ 1- \frac{r^{2n} }{(r^2+a_0^2)^n} \right] \pd_u^2 B_0 \,.
\end{align}
\end{subequations}
The integration of the $J_s^{(n=0)}$ piece is exactly the same as \cite{Bombini:2017sge}, giving
\be
\begin{split}
  B^{(n=0)} (z, \bar z)  &= \frac{b^2}{\pi a_0^2} \left[  - \half \, \hat D_{2222} + \frac{1}{|1-z|^4} \, \left( 2(1  + |z|^2) \hat D_{3311} -\pi \right)  \right] \\
&=  \frac{b^2}{\pi a_0^2} \ \pd \bar \pd \left[ - \frac{\pi}{2}\, \frac{1}{|1-z|^2} + |z|^2 \hat D_{1122}   \right]  ,
\end{split}
\ee
while the $n>0$ term can be integrated noticing that it can be written as 
\be
J_s^{(n>0)} = - \frac{R^2}{2a_0^2} \sum_{p=1}^n {n \choose k} (-1)^{p} (B_+ B_-)^{p+1} \pd_u^2 B_0 \,,
\ee
so that, using the fact that $\pd_{u'} = - \pd_u $ by the dependence upon $\tau'_E-\tau_E, \sigma'- \sigma$ of the integrand, we can integrate by parts each $p$-term as  
\be
\begin{split}
  |z|^2 I_p (z, \bar z)  &= \pd_u^2 \int d^3 r_e' \sqrt{-g} \, B_0 ( {\bf r}_e'| 0,0 ) B_+^{p+1}({\bf r}')  B_-^{p+1}({\bf r}') B_0 ( {\bf r}_e'| t_e, y) \\
&= 4 (\bar z^2 \, \bar \pd^2 + \bar z \, \bar \pd ) \left( |z|^2 \hat{D}_{(p+1)(p+1)22} \right) , 
\end{split}
\ee  
where we have used the notation of app.~\ref{sec:AppB}, \ref{sec:AppC},
and that $\pd_u^2 = - (\bar z \, \bar \pd )^2 =- \bar z^2 \, \bar \pd^2  - \bar z \, \bar \pd$. Summing all the $p$ terms, we get the final result
\begin{subequations}\label{3charge_HHLL}
\begin{align}
{\cal C}_{(1,0,n)}^{{\cal O} (b^2)}  &= {\cal C}_{n=0} + {\cal C}_{n>0}\,, \\
{\cal C}_{n=0} &= + \frac{b^2}{\pi a_0^2}\left[ \frac{1}{|1-z|^4} \, \left( 2(1  + |z|^2) \hat D_{3311} -\pi \right)  - \half \, \hat D_{2222}  \right] ,\label{3charge_HHLL_1} \\
{\cal C}_{n>0}&= - \frac{b^2}{\pi a_0^2} \sum_{p=1}^n {n \choose k} (-1)^{p} \frac{1}{z} (\bar z \, \bar \pd^2 + \bar \pd) \left( |z|^2 \hat{D}_{(p+1)(p+1)22} \right) .\label{3charge_HHLL_2}
\end{align}
\end{subequations}
Adding the free contribution, we have the first four-point functions involving $\frac{1}{8}$-BPS operators in AdS$_3$: 
\begin{equation}
\label{eq:HHLL1}
    \begin{split}
        {\cal C}_{(1,0,n)} = \frac{1}{|1-z|^4}+ \frac{b^2}{\pi a_0^2} & \Big[ \frac{1}{|1-z|^4} \, \left( 2(1  + |z|^2) \hat D_{3311} -\pi \right)  - \half \, \hat D_{2222} \\
        & \quad -  \sum_{p=1}^n {n \choose k} (-1)^{p} \frac{1}{z} (\bar z \, \bar \pd^2 + \bar \pd) \left( |z|^2 \hat{D}_{(p+1)(p+1)22} \right) \Big] . 
    \end{split}
\end{equation}

\subsection{HHLL 4-point functions in the $(k,m,n)=(m,m,0)$ geometry}
\label{subsec:110}

We want now to study the bosonic four-point function in a different three-charge geometry, the $k=m$, $n=0$ one; this is a superdescendant of a two charge geometry and was built in \cite{Giusto:2013bda, Bena:2016ypk, Bena:2017xbt, Bakhshaei:2018vux}:
\begin{subequations}
    \begin{align}
        ds_6^2 &= - \frac{2}{\sqrt{\mathcal{P}}} (dv+\beta) \left[ du + \omega   + \half \,  \F (dv+\beta) \right] + \sqrt{\mathcal{P}} \, ds_4^2 \, ,  \\
        ds_4^2 &= \Sigma \left( \frac{dr^2}{r^2 +a^2} + d \theta^2 \right) + (r^2+a^2)\sin^2 \theta d\phi^2 + r^2 \cos^2 \theta \, d\psi^2 \,, \\
        Z_1 &=  \frac{R^2}{2 Q_5} \, \frac{1}{\Sigma} \left[ \left( 2a^2 +b^2  \right) + \frac{a^{2{m}} \cos^{2{m}} \theta}{(r^2+a^2)^{m}}\, b^2 \cos \left(2 m \hat v \right) \right]  \,, \quad Z_2 = \frac{Q_5}{\Sigma} \,, \\
        Z_4 &= \frac{R}{\Sigma} \,b \, \frac{a^{m} \cos^{m} \theta}{(r^2+a^2)^{\frac{m}{2}}} \,  \cos \left( m \hat v \right) \,, \qquad \Sigma = r^2 + a^2 \cos^2 \theta \,, \\
         {\cal F} &= - \frac{ b^2 }{r^2 + a^2 \sin^2 \theta}  \left( 1- \frac{a^{2m} \cos^{2m} \theta}{(r^2+a^2)^{m}} \right) \,, \qquad {\cal P} = Z_1 Z_2 - Z_4^2  \,, \\
        \omega &= \omega_0 - \frac{R}{\sqrt{2} \,\Sigma} \, {\cal F}  \left[ (r^2+a^2)\sin^2 \theta \, d \phi +  r^2 \cos^2 \theta \, d\psi  \right] \,, \\
        \beta_0 &= \frac{R\, a^2}{\sqrt{2} \, \Sigma} ( \sin^2 \theta \, d\phi - \cos^2 \theta \, d\psi), \quad \omega_0 = \frac{R\, a^2}{\sqrt{2} \, \Sigma} ( \sin^2 \theta \, d\phi + \cos^2 \theta \, d\psi),
    \end{align}
\end{subequations}
where $\hat v = \frac{\sqrt{2}\, v}{R} - \psi$. As before, it may be useful to rewrite this geometry as 
\begin{equation}
    ds_6^2 = V^{-2} g_{\mu\nu} dx^\mu dx^\nu + G_{ab} (d\theta^a + A^a{}_\mu dx^\mu) (d\theta^b + A^b{}_\nu dx^\nu), 
\end{equation}
where $V^2 = \frac{\det G}{\det G_0} $, with $G_0$ metric of the round $S^3$. For sake of semplicity, we report here only the $g_3$ for the case $k=m=1$, that is 
\begin{equation}\label{eq_3Dmetric110}
    \frac{ds_3^2}{\sqrt{Q_1 Q_5}}  = - \left( 1 + \frac{r^2 - b^2}{a_0^2} - \frac{b^4}{4a_0^4} \right) d\tau^2 + \frac{r^2}{a_0^2}\, d\sigma^2 + \frac{r^2 + a_0^2 \left(1 - \frac{b^2}{2a_0^2}\right)}{\left( r^2 +a_0^2 - \frac{b^2}{2} \right)^2} \, dr^2 \,, 
\end{equation}
since the $m>1$ differs from this only via higher scalar harmonics term, that gives only terms are integrated out in the extraction of the solution.

We will again perform the same procedure, in order to compute the four-point function; this will give us the source, once we have integrated on the appropriate harmonic, that is 
\begin{equation}\label{eq:sourcemm}
    J_s = - \frac{b^2}{2a_0^2} \left[ 2 a_0^2 \, \frac{r}{r^2+a_0^2} \, \pd_r B_0 - 2 \frac{a_0^4}{(r^2+a_0^2)^2} \, \pd_{\tau}^2 B_0   \right] ,
\end{equation}
for all $m$. This is due to the fact that $m$ sources higher harmonics that are projected out. This source can be easily integrated and, using the notation of App.~E of \cite{Bombini:2017sge} that we will briefly review in App.~\ref{sec:AppC}, we get
\begin{equation}\label{eq:BulkIntC110}
    {\cal C}_{(m,m,0)}^{{\cal O} (b^2)} = \frac{b^2}{\pi a_0^2} \left[  \frac{I_1 + I_2}{2} -  I_3 \right] ,
\end{equation}
that is, explicitly,
\begin{equation}
    {\cal C}_{(m,m,0)}^{{\cal O} (b^2)}  =\frac{b^2}{\pi a_0^2} \left[ \frac{1}{|1-z|^4} \left( 2(1+ |z|^2) \hat{D}_{3311} -\pi \right) - \half \, \hat{D}_{2222} \right] .
\end{equation}
This means that the full correlator is
\begin{equation}\label{eq:4pf110}
    {\cal C}_{(m,m,0)} =\frac{1}{|1-z|^4} + \frac{b^2}{\pi a_0^2} \left[ \frac{1}{|1-z|^4} \left( 2(1+ |z|^2) \hat{D}_{3311} -\pi \right) - \half \, \hat{D}_{2222} \right] .
\end{equation}

\subsection{HHLL 4-point functions in the $(k,m,n,q)=(2,0,0,1)$ geometry}
The geometry with $k=2,m=n=0, q=1$, built in \cite{Ceplak:2018pws},  dual to the R-state obtained via spectral flow from the NS-state
\begin{equation}
    |2,0,0,1\ket = \left[ |0\ket \right]^{N_a} \left[ \left(  G_{-\half}^{+1}  G_{-\half}^{+2} + \frac{1}{k} \, J_0^3 L_{-1} \right) \left| O^{--} \right\ket_2 \right]^{N_b} \,,
\end{equation}
with the regularity $N_a + 2 N_b = N$, has the usual form \eqref{eq:FullMetrickmn} with 
\begin{subequations}
\begin{align}
    ds_6^2 &= - \frac{2}{\sqrt{\mathcal{P}}} (dv+\beta) \left[ du + \omega   + \half \,  \F (dv+\beta) \right] + \sqrt{\mathcal{P}} \, ds_4^2 \, ,  \\
    ds_4^2 &= \Sigma \left( \frac{dr^2}{r^2 +a^2} + d \theta^2 \right) + (r^2+a^2)\sin^2 \theta \, d\phi^2 + r^2 \cos^2 \theta \, d\psi^2 \,, \\
    Z_1 &= \frac{Q_1}{\Sigma} \,, \quad Z_2 = \frac{Q_5}{\Sigma} \,, \quad Z_4 = 0 \,, \quad {\cal P} = Z_1 Z_2 \,, \\
    {\cal F} &=-  \frac{b^2 \left(3 r^4 + 8  r^2 a^2 + 5 a^4 - a^2 \left(r^2 + 2 a^2\right) \sin ^2 \theta \right)}{3 \left(r^2+a^2\right)^3} \,, \\
    \omega &= \omega_0 + \frac{1}{6} \, \frac{R \, b^2}{\sqrt{2} \, \Sigma} \, \frac{(3 r^4 + 8 r^2 a^2 + 6 a^4) \sin^2 \theta \, d\phi + r^2 (3r^2+4a^2)\cos^2 \theta\, d\psi }{(r^2+a^2)^2} \,, 
\end{align}
\end{subequations}
where $\omega_0 = \frac{R\, a^2}{\sqrt{2} \, \Sigma} \, (\sin^2 \theta \, d\phi + \cos^2 \theta \, d\psi)$, and with the regularity
\begin{equation}
    a^2 + \frac{b^2}{2} = \frac{Q_1 Q_2}{R^2} \equiv a_0^2 \,.
\end{equation}
This peculiar metric is separable at all order in $b$, and it is easy to see that, performing the same small $b$ procedure, we will get a source that is
\begin{equation}
    \begin{split}
    J_s = \frac{b^2}{2a_0} & \, \Bigg[ - 2 a_0^2 \, \frac{r}{r^2+a_0^2} \, \pd_r B_0 + \frac{1}{3} \, \left( 2 \left(\frac{a_0^2}{r^2+a_0^2}\right)^3   +9 \left(\frac{a_0^2}{r^2+a_0^2}\right)^2 \right) \pd_\tau^2 B_0 \\
    & \quad - \left( \left(\frac{a_0^2}{r^2+a_0^2}\right)^3 + \left(\frac{a_0^2}{r^2+a_0^2}\right)^2 \right) \pd_\tau \pd_\sigma B_0 + \frac{1}{3} \left(\frac{a_0^2}{r^2+a_0^2}\right)^3 \pd_\sigma^2 B_0 \Bigg] . 
    \end{split}
\end{equation}
This can be easily integrated noticing that
\begin{equation}
    (B_+ B_-)^2 = \left(\frac{a_0^2}{r^2+a_0^2}\right)^2 \,, \quad (B_+ B_-)^3 = \left(\frac{a_0^2}{r^2+a_0^2}\right)^3 \,,
\end{equation}
and that
\begin{equation}
     - 2 a_0^2 \, \frac{r}{r^2+a_0^2} \, \pd_r B_0 
\end{equation}
is the term we already encountered in eq.~\eqref{eq:sourcemm}, so that, using the results of app.~\ref{sec:AppB} and taking into account the map from the cylinder to the plane and the free part, we get
\begin{equation}\label{eq:4pfq1}
    \begin{split}
        {\cal C}_{(2,0,0,1)}  (z, \bar z) = \frac{1}{|1-z|^4} - \frac{b^2}{2 \pi a_0}  \, \frac{1}{|z|^2} \, & \Bigg[ |z|^2 \hat{D}_{2222} - \frac{1}{3} (z \pd + \bar z \bar \pd)^2 \left( 2 |z|^2 \hat{D}_{3322} + 9 |z|^2 \hat{D}_{2222}  \right) \\
        & \qquad + (z \pd + \bar z \bar \pd) (z \pd - \bar z \bar \pd) \left(  |z|^2 \hat{D}_{3322} + |z|^2 \hat{D}_{2222}  \right) \\
        & \qquad \quad - \frac{1}{3} (z \pd - \bar z \bar \pd)^2 \left( |z|^2 \hat{D}_{3322} \right) \Bigg].
    \end{split}
\end{equation}
With this correlator, we have computed all the four-point functions involving $\frac{1}{8}$-BPS operators whose dual geometries are explicitly known that we have described in sec.~\ref{sec:CFT}. 
Using the properties of the D-functions reported in app.~\ref{sec:AppC}, it is straightforward to notice that all the four-point function results (\ref{eq:HHLL1}, \ref{eq:4pf110}, \ref{eq:4pfq1}) posses the symmetry under the exchange of the two bosonic operators $z_3 \leftrightarrow z_4$, i.e. $z \to z^{-1}$, as required. This is a first check on the validity of the computations. 

\section{From HHLL to LLLL correlators}
\label{sec:LLLL}
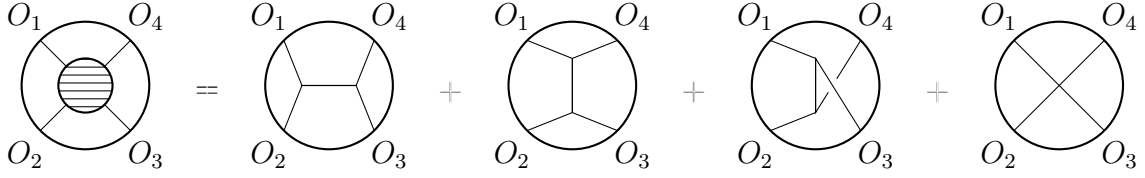
\begin{figure}   
    \centering
    \begin{tikzpicture}[scale=0.6]
    \draw (1,1) node[above]{$\phantom{1} \hspace{2mm} O_4$} --(0,0) -- (-1,-1) node[below]{$O_2 \hspace{2mm} \phantom{1}$};
    \draw (1,-1) node[below]{$\phantom{4} \hspace{2mm} O_3$}  --(0,0) -- (-1,+1) node[above]{$O_1 \hspace{2mm} \phantom{1}$};
    \draw [thick] (0,0) circle [radius=1.414];
    \draw [fill, white] (0,0) circle [radius=0.6]; 
    \draw [pattern= horizontal lines, thick] (0,0) circle [radius=0.6]; 
    \end{tikzpicture}
    \centering
    \begin{tikzpicture}[scale=0.6] 
    \draw [draw=white, thin] (0,-0.01)-- (0,0) node[baseline]{$=$}  -- (0,+0.01);
    \draw [draw=white, thin] (0,1.9)-- (0,-1.9);
    \draw [draw=white, thin] (0.4, 0) --(-0.4,0);
    \end{tikzpicture}
    \centering
    \begin{tikzpicture}[scale=0.6] 
    \draw  (1,-1) node[below]{$\phantom{4} \hspace{2mm} O_3$} -- (+0.6,0) -- (-0.6,0) -- (+0.6,0)  -- (1,1) node[above]{$\phantom{1} \hspace{2mm} O_4$} ;
    \draw (-1,+1) node[above]{$O_1 \hspace{2mm} \phantom{1}$}   -- (-0.6,0) -- (+0.6,0) -- (-0.6,0) --   (-1,-1) node[below]{$O_2 \hspace{2mm} \phantom{1}$};
    \draw [thick] (0,0) circle [radius=1.414];
    \end{tikzpicture}
    \centering
    \begin{tikzpicture}[scale=0.6] 
    \draw [draw=white, thin] (0,-0.01)-- (0,0) node[baseline]{$+$}  -- (0,+0.01);
    \draw [draw=white, thin] (0,1.9)-- (0,-1.9);
    \draw [draw=white, thin] (0.25, 0)--(-0.25,0);
    \end{tikzpicture}
    \centering
    \begin{tikzpicture}[scale=0.6] 
    \draw (1,1) node[above]{$\phantom{1} \hspace{2mm} O_4$} -- (0,+0.6) -- (0,-0.6) -- (-1,-1) node[below]{$O_2 \hspace{2mm} \phantom{1}$};
    \draw (1,-1) node[below]{$\phantom{4} \hspace{2mm} O_3$}  -- (0,-0.6) -- (0,+0.6) -- (-1,+1) node[above]{$O_1 \hspace{2mm} \phantom{1}$};
    \draw [thick] (0,0) circle [radius=1.414];
    \end{tikzpicture}
    \centering
    \begin{tikzpicture}[scale=0.6] 
    \draw [draw=white, thin] (0,-0.01)-- (0,0) node[baseline]{$+$}  -- (0,+0.01);
    \draw [draw=white, thin] (0,1.9)-- (0,-1.9);
    \draw [draw=white, thin] (0.25, 0)--(-0.25,0);
    \end{tikzpicture}
     \centering
    \begin{tikzpicture}[scale=0.6] 
    \draw (1,1) node[above]{$\phantom{1} \hspace{2mm} O_4$}   -- (0,-0.6) -- (0,+0.6) -- (-1,+1) node[above]{$O_1 \hspace{2mm} \phantom{1}$};
    \draw [white, thick] (0.46875 , 0.15) -- (0.28125, -0.15);  
     \draw  (1,-1) node[below]{$\phantom{4} \hspace{2mm} O_3$} -- (0,+0.6) -- (0,-0.6) -- (-1,-1) node[below]{$O_2 \hspace{2mm} \phantom{1}$};
    \draw [thick] (0,0) circle [radius=1.414];
    \end{tikzpicture}
    \centering
    \begin{tikzpicture}[scale=0.6] 
    \draw [draw=white, thin] (0,-0.01)-- (0,0) node[baseline]{$+$}  -- (0,+0.01);
    \draw [draw=white, thin] (0,1.9)-- (0,-1.9);
    \draw [draw=white, thin] (0.25, 0)--(-0.25,0);
    \end{tikzpicture}
    \centering
    \begin{tikzpicture}[scale=0.6] 
    \draw (1,1) node[above]{$\phantom{1} \hspace{2mm} O_4$} --(0,0) -- (-1,-1) node[below]{$O_2 \hspace{2mm} \phantom{1}$};
    \draw (1,-1) node[below]{$\phantom{4} \hspace{2mm} O_3$}  --(0,0) -- (-1,+1) node[above]{$O_1 \hspace{2mm} \phantom{1}$};
    \draw [thick] (0,0) circle [radius=1.414];
    \end{tikzpicture}
    \caption{In this figure we report pictorially all the channels that enters in the computation of a generic four-point function. Recall that each diagram  correspond to a channel contribution to the correlator ${\cal C}= \langle O_1(z_1,\bar{z}_1)O_2(z_2,\bar{z}_2)O_3(z_3,\bar{z}_3)O_4(z_4,\bar{z}_4)\rangle$. }
    \label{fig:allchannelWD}
\end{figure}

The purpose of this section is to follow the arguments of \cite{Giusto:2018ovt} to extract the s-channel of a correlator containing only single-trace operators from the correlators we computed in the previous sections. Here we focus on four-point functions of the type
\begin{equation}
    \label{eq:LLLL}
    \langle O_1(z_1,\bar{z}_1)O_2(z_2,\bar{z}_2)O_3(z_3,\bar{z}_3)O_4(z_4,\bar{z}_4)\rangle=\frac{1}{|z_{12}|^{2\Delta_1}|z_{34}|^{2\Delta_3}} \, \mathcal{G}(z,\bar{z}) ,
\end{equation}
where we assume that the conformal dimensions satisfy $\Delta_1=\Delta_2$ and $\Delta_3=\Delta_4$ and they does not scale with the central charge. From the general structure of the correlator we can write it as a sum over channels\footnote{Here we use the usual notation for the $s$, $t$ and $u$ channels used when computations with Witten diagrams are performed. }
\begin{equation}\label{eq:allG}
 \mathcal{G}=\mathcal{G}_s+\mathcal{G}_t+\mathcal{G}_u+\mathcal{G}_{\text{cont}} \,,
\end{equation}
where the $s,t,u$ contributions take into account the single-trace operators exchanged in each of those channel, while $\mathcal{G}_{\text{cont}}$ encode the contribution of the possible contact terms. On the complex plane, the three $s,t,u$ channels  are $z\to 0$, $z\to 1$, $z\to \infty$, respectively. Each of these terms has a bulk picture in terms of Witten diagram. In particular we have that the $s,t,u$ contributions come from three different Witten diagrams arising from three-point vertexes in the bulk and the contact terms come from the four-point vertexes. A pictorial representation of eq. \eqref{eq:allG} is reported in fig.~\ref{fig:allchannelWD}. 

In the case of Heavy-Light (HHLL) four-point functions we bypassed the Witten diagram machinery and we computed them as a two-point function of the light operators in a non-trivial background, sourced by the heavy operators. The possibility to extract the all-light (LLLL) four-point functions from the HHLL relies in the fact that the heavy multi-trace operators involved in the HHLL correlators are made of the single-trace constituents involved in the LLLL one. The number of these constituents is controlled by the free parameter that we called $b$ in the previous sections.

\begin{figure} 
    \centering
    \begin{tikzpicture}[scale=1]
    \draw [thick] (0,0) circle [radius=1.414];
    \draw  (-1,+1) node[above]{$O_L \hspace{2mm} \phantom{1}$} -- (-0.4, 0) -- (-1,-1) node[below]{$\bar{O}_L \hspace{2mm} \phantom{1}$};
    \draw [photon, thick] (-0.4, 0) -- (+0.4, 0);
    \draw [ultra thick, blue] (0.6, 0) circle [radius=0.2];
    \draw [blue, thick] (0.4586, -0.1414) -- (0.6, 0) node[below=6]{$ \phantom{1} \vspace{20mm} O_H$} -- (0.7414, 0.1414);
    \draw [blue, thick] (0.7414, -0.1414) -- (0.6, 0) -- (0.4586, 0.1414);
    \end{tikzpicture}
    \begin{tikzpicture}[scale=1] 
    \draw [draw=black, thick, ->] (-0.5, 0)-- (0,0) node[above]{$\frac{b^2}{2a_0^2} \to \frac{1}{N}$}  -- (+0.5, 0);
    \draw [draw=white, thin] (0,1.75) -- (0,-1.75);
    \draw [draw=white, thin] (0.4, 0) --(-0.4,0);
    \end{tikzpicture}
    \centering
    \begin{tikzpicture}[scale=1] 
    \draw [red] (1,-1) node[below]{$\phantom{1} \hspace{5.5mm} \textcolor{red}{O^{--}}$} -- (+0.6,0)  -- (1,1) node[above]{$\phantom{1} \hspace{5.5mm} \textcolor{red}{O^{++}}$} ;
    \draw (-1,+1) node[above]{$O_L \hspace{2mm} \phantom{1}$}   -- (-0.6,0) -- (+0.6,0) -- (-0.6,0) --   (-1,-1) node[below]{$\bar{O}_L \hspace{2mm} \phantom{1}$};
    \draw [thick] (0,0) circle [radius=1.414];
    \end{tikzpicture}
    \caption{The spectral flow from the Ramond sector to the Neveu-Schwarz sector ``lightens'' the heavy operator, since the $\left| ++ \right\rangle_1$ states flow into the NS-vacua; since we are keeping $\frac{b^2}{2a_0^2}$ small, it is like having $N^{++} \ll N$. Taking then the limit $\frac{b^2}{2a_0^2} \to \frac{1}{N}$ while in the NS sector, we obtain a LLLL four-point function s-channel out of the spectrally-flowed HHLL one.    }
    \label{fig:HHLLtoLLLLs}
\end{figure}
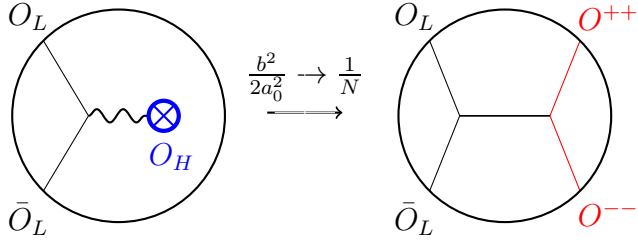

Since the D1D5 has two sector, as explained in sec.~\ref{sec:CFT}, we can use the spectral flow transformation to flow from the NS to the R sector, and vice versa. In particular, the HHLL correlators described above are computed in the R sector, and would be interesting to see how those flow into the NS sector. This interest arise in the fact that the spectral flow, that acts on the generator as \cite{Avery:2010qw}
\begin{equation}
    L_n \mapsto L_n + J^3_m + \frac{1}{4} \, \delta_{m,0} \,, \quad J^3_m \mapsto J_m^3 - \half \, \delta_{m,0} \,, \quad J^m_m \mapsto J^\pm_{m\mp 1} \,, \quad G^{\pm,  A}_m \mapsto G^{\pm , A}_{m \pm \half} \,,
\end{equation}
on the R-vacua acts as 
\begin{equation}
    \left | ++ \right \rangle_1 \mapsto |0\ket_1 \,, \quad |00\ket_1 \mapsto O^{--} |0\ket_1 \,,
\end{equation}
where $O^{--}$, often dubbed as fermionic operator\footnote{The nomenclature comes from the fact that it is a product of two fermions of the two different chiral sectors of the theory, i.e. $\psi$ and $\tilde \psi$. } \cite{Giusto:2015dfa, Galliani:2016cai, Galliani:2017jlg, Bombini:2017sge}, is one of the element of the family
\begin{equation}\label{eq:OFfamily}
    O^{\alpha \dot \alpha} (z, \bar z) = \sum_{r=1}^N \frac{-i \ep_{\dot A \dot B}}{\sqrt{2N}} \, \psi^{\alpha \dot A}_{(r)} \tilde{\psi}^{\dot\alpha  \dot B}_{(r)} \,.
\end{equation}
This last point is crucial, since it means that the heavy state \eqref{eq:3ch} flows in the NS sector to
\begin{equation}\label{eq:kmnSF}
    |k,m,n\ket \mapsto \left[|0\ket_1 \right]^{N_1} \prod_{k,m,n}  \left[ \frac{(L_{-1})^n}{n!} \, \frac{(J_0^+)^m}{m!} \, O^{--} |0\ket_1  \right]^{N_{k,m,n}} \,.
\end{equation}
This explicitly means that now we can play with the number of operators insertion by controlling $N_{k,m,n}/N$, i.e. $\frac{b^2}{2a_0^2}$; in the NS the $b^2 \ll 2 a_0^2$ limit is the ``lightening'' limit $N^{k,m,n} \ll N$; sending
\begin{equation}
    \frac{b^2}{2a_0^2} = \frac{N_{k,m,n}}{N} \to \frac{1}{N} \,,
\end{equation}
we reduce the HHLL to a LLLL in the $s$-channel \cite{Giusto:2018ovt}. One may wonder why this limit does not reproduce the full LLLL correlator, but only the $s$-channel contribution; this is due to the fact that, in the HHLL computation in the R-sector, no single-trace operator is exchanged in the channels where one heavy and one light operator fuse together, i.e. in the $t$- or $u$-channels; since this statement is true for all values of $b$, this survives the ``lightening'' limit. This implies that the correlator we extract via this ``lightening'' ansatz of \cite{Giusto:2018ovt}, lacks of the $t$- and $u$-channel contribution, and the contact terms (since even in contact terms heavy and light operator fuse, so are not taken in account). A pictorial representation of that is reported in fig.~\ref{fig:HHLLtoLLLLs}. In \cite{Giusto:2018ovt} it is also shown how, at least for a certain subset of all the possible D1D5 four-point functions, it is possible to fix unambiguously all the missing terms, effectively reconstructing the full LLLL four-point functions. In particular, they focus on  four-point functions where the four operators are all the possible combinations of the $O^{\alpha \dot \alpha}$ operators of eq.~\eqref{eq:OFfamily}. 

We start from the multi-particle heavy states that create the background in the Ramond sector. As explained in eq.~\eqref{eq:kmnSF}, these heavy operators flow in the NS in:
\begin{subequations}
\begin{align}
|1,0,n\rangle &\to \left(L_{-1}^n O^{--}\right)^{N_b}\label{q1}  \,,\\
|m,m,0\rangle &\to \left((J^+_{0})^m O^{--}\right)^{N_b}\label{q2} \,,
\end{align}
\end{subequations}
with $N_b=Nb^2/(2a_0^2)$ in the parameter controlling the number of single trace operator inside the heavy state. 

Let us start analyzing the LLLL four-point function \eqref{eq:LLLL}, coming from the states in \eqref{q1}
\begin{equation}
    O_1=\left(L_{-1}^nO^{--}\right),\quad O_2=\left(L_{+1}^nO^{++}\right),\quad O_3=O_{\text{bos}},\quad O_4=\bar{O}_{\text{bos}} \,.
\end{equation}
Following \cite{Giusto:2018ovt} we have that the s-channel of the this correlator is given by the correlator \eqref{eq:HHLL1} flowed in the NS sector and setting $N_b=1$:
\begin{equation}
\label{LLLL1}
    \begin{split}
    \mathcal{G}_s^{(1,0,n)}(z,\bar{z})=\frac{|1-z|^4}{\pi N} & \Bigg[ \frac{1}{|1-z|^4} \, \left( 4(1  + |z|^2) \hat D_{3311} -2\pi \right)  - \, \hat D_{2222} \\
        & \quad -  2\sum_{p=1}^n {n \choose k} (-1)^{p} \frac{1}{z} (\bar z \, \bar \pd^2 + \bar \pd) \left( |z|^2 \hat{D}_{(p+1)(p+1)22} \right) \Bigg] ,
\end{split}
\end{equation}
i.e., for $n=1$ case, 
\begin{equation}
    \begin{split}
        \begin{tikzpicture}[scale=0.6, baseline={([yshift=-.5ex]current bounding box.center)},vertex/.style={anchor=base,
            circle}]] 
        \draw  (1,-1) node[below]{$\phantom{1} \hspace{10mm}   L_{+1} O^{++}$} -- (+0.6,0) -- (-0.6,0) -- (+0.6,0)  -- (1,1) node[above]{$\phantom{1} \hspace{10mm}   L_{-1} O^{--}$} ;
        \draw (-1,+1) node[above]{$ O_B \hspace{2mm} \phantom{1}$}   -- (-0.6,0) -- (+0.6,0) -- (-0.6,0) --   (-1,-1) node[below]{$\bar{O}_B \hspace{2mm} \phantom{1}$};
        \draw [thick] (0,0) circle [radius=1.414];
        \end{tikzpicture} %
    =   \frac{1}{\pi N} & \left[  \frac{ 4(1  + |z|^2) \hat D_{3311} -2\pi }{|1-z|^4}  -   \hat D_{2222}  + 2 \,   \frac{(\bar z \bar \pd^2 + \bar \pd)}{z} \left( |z|^2 \hat D_{2222} \right)      \right] .
    \end{split}
\end{equation}
Notice that taking the case $n=0$ we obtain the s-channel for the LLLL correlators with 
\begin{equation}
    O_1=O^{--} \, ,\quad O_2=O^{++} \,,\quad O_3=O_{\text{bos}},\quad O_4=\bar{O}_{\text{bos}} \,.
\end{equation}
This contribution could be directly found by performing this analysis starting from the HHLL correlator computed in \cite{Bombini:2017sge}, and applying the prescription described above. The results agree and they are given by
\begin{equation}
\label{LLLL2}
    \begin{split}
    \mathcal{G}_s(z,\bar{z})=\frac{|1-z|^4}{\pi N} & \left[ \frac{1}{|1-z|^4} \, \left( 4(1  + |z|^2) \hat D_{3311} -2\pi \right)  - \, \hat D_{2222} \right] ,
\end{split}
\end{equation}
i.e. 
\begin{equation}
    \begin{tikzpicture}[scale=0.6, baseline={([yshift=-.5ex]current bounding box.center)},vertex/.style={anchor=base,
    circle}]] 
    \draw  (1,-1) node[below]{$\phantom{1} \hspace{7mm} O^{++}$} -- (+0.6,0) -- (-0.6,0) -- (+0.6,0)  -- (1,1) node[above]{$\phantom{1} \hspace{7mm} O^{--}$} ;
    \draw (-1,+1) node[above]{$ O_B \hspace{2mm} \phantom{1}$}   -- (-0.6,0) -- (+0.6,0) -- (-0.6,0) --   (-1,-1) node[below]{$\bar{O}_B \hspace{2mm} \phantom{1}$};
    \draw [thick] (0,0) circle [radius=1.414];
    \end{tikzpicture} %
    =  \frac{1}{\pi N}\left[ \frac{1}{|1-z|^4} \, \left( 4(1  + |z|^2) \hat D_{3311} -2\pi \right) -  \hat D_{2222} \right] .
\end{equation}
With similar analysis we can extract the s-channel for the LLLL correlator \eqref{eq:LLLL} with 
\begin{equation}
    O_1=\left((J_{0}^+)^mO^{--}\right),\quad O_2=\left((J_{0}^-)^mO^{++}\right),\quad O_3=O_{\text{bos}},\quad O_4=\bar{O}_{\text{bos}},
\end{equation}
coming from the heavy states \eqref{q2}. The result for the s-channel reads
\begin{equation}
\label{LLLL3}
    \begin{split}
    \mathcal{G}^{(m,m,0)}_s(z,\bar{z})=\frac{|1-z|^4}{\pi N} & \left[ \frac{1}{|1-z|^4} \, \left( 4(1  + |z|^2) \hat D_{3311} -2\pi \right)  - \, \hat D_{2222}\right] ,
\end{split}
\end{equation}
i.e. 
\begin{equation}
    \begin{tikzpicture}[scale=0.6, baseline={([yshift=-.5ex]current bounding box.center)},vertex/.style={anchor=base,
    circle}]] 
    \draw  (1,-1) node[below]{$\phantom{1} \hspace{10mm} J_0^- O^{++}$} -- (+0.6,0) -- (-0.6,0) -- (+0.6,0)  -- (1,1) node[above]{$\phantom{1} \hspace{10mm} J_0^+ O^{--}$} ;
    \draw (-1,+1) node[above]{$ O_B \hspace{2mm} \phantom{1}$}   -- (-0.6,0) -- (+0.6,0) -- (-0.6,0) --   (-1,-1) node[below]{$\bar{O}_B \hspace{2mm} \phantom{1}$};
    \draw [thick] (0,0) circle [radius=1.414];
    \end{tikzpicture} %
    =  \frac{1}{\pi N}\left[ \frac{1}{|1-z|^4} \, \left( 4(1  + |z|^2) \hat D_{3311} -2\pi \right) -  \hat D_{2222} \right] .
\end{equation}

For sake of completeness, we report here also a result already obtained in \cite{Galliani:2017jlg, Giusto:2018ovt}, where it was computed the HHLL four-point function with four fermionic operators by which it is possible to extract the LLLL s-channel contribution
\begin{equation}
   \left[ \bra O^{--} (0) O^{++} (\infty) O^{++} (1) O^{--} (z, \bar z) \ket \right]_{\rm s-channel} \equiv \frac{1}{|1-z|^2} \, {\cal G}_s^{\rm fer} (z, \bar z) , 
\end{equation}
that is
\begin{equation}\label{eq:FERMres}
    {\cal G}_s^{\rm fer} = \frac{1}{\pi \, N} \left[ 2\, |z|^2 |1-z|^2 \hat{D}_{1122} - \pi  \right] , 
\end{equation}
i.e.
\begin{equation}  
    \begin{tikzpicture}[scale=0.6, baseline={([yshift=-.5ex]current bounding box.center)},vertex/.style={anchor=base,
    circle}]] 
    \draw  (1,-1) node[below]{$\phantom{1} \hspace{7mm} O^{++}$} -- (+0.6,0) -- (-0.6,0) -- (+0.6,0)  -- (1,1) node[above]{$\phantom{1} \hspace{7mm} O^{--}$} ;
    \draw (-1,+1) node[above]{$ O^{--} \hspace{7mm} \phantom{1}$}   -- (-0.6,0) -- (+0.6,0) -- (-0.6,0) --   (-1,-1) node[below]{$O^{++} \hspace{7mm} \phantom{1}$};
    \draw [thick] (0,0) circle [radius=1.414];
    \end{tikzpicture} %
    =  \frac{1}{\pi N} \left( 2 |z|^2  \hat{D}_{1122} - \frac{\pi}{|1-z|^2}     \right) .
\end{equation}

In the next section we will explain how, using Ward identities, we will be able to reconstruct the full LLLL correlators out of the one written here.

\section{Ward Identities for four-point functions}
\label{sec:WI}

As explained in sec.~\ref{sec:CFT}, the D1D5 CFT has a ${\cal N}=(4,4)$ superconformal algebra with a chiral $SU(2)$ Kac-Moody symmetry, plus an $SU(2) \times SU(2)$ custodial global symmetry.

The main purpose of this section is then to find the Ward Identities (WI) for the generators $L_{-1},\,J^a_{0}, \, G^{+ A}_0$ appearing in the correlators computed in the previous section. Concretely, one wants to find a differential operators, encoding the effects of the insertion of the algebra generators, that acts on the correlators without any insertion, namely the four-point function in \eqref{LLLL2}.

The WI referring to the spectrally-flowed state (\ref{q1}) can be found by considering the following correlator
\begin{equation}
\label{eq:WI1}
\begin{split}
\langle\left(L_{-1}^nO^{--}(z_1,\bar{z}_1)\right)\left(L_{+1}^nO^{++}(z_2,\bar{z}_2)\right)O_{\text{bos}}(z_3,\bar{z}_3)\bar{O}_{\text{bos}}(z_4,\bar{z}_4)\rangle , 
\end{split}
\end{equation}
and finding the differential representation of the operators $L_{\pm 1}^n$ on the operators. In order to do that we focus on the case $n=1$ for simplicity. Using \eqref{eq:exp} we have
\begin{equation}
\label{eq:WI2}
\begin{split}
&\langle\left(L_{-1}O^{--}(z_1,\bar{z}_1)\right)\left(L_{+1}O^{++}(z_2,\bar{z}_2)\right)O_{\text{bos}}(z_3,\bar{z}_3)\bar{O}_{\text{bos}}(z_4,\bar{z}_4)\rangle\\
&=\oint_{z_1}\frac{dw_1}{2\pi i}\oint_{z_2}\frac{dw_2}{2\pi i}w_2^2\langle T(w_1)O^{--}(z_1,\bar{z}_1)T(w_2)O^{++}(z_2,\bar{z}_2)O_{\text{bos}}(z_3,\bar{z}_3)\bar{O}_{\text{bos}}(z_4,\bar{z}_4)\rangle\\
&=\left[\left( z_2^2 \pd_{z_2} + 2h_1 z_2  \right)\pd_{z_1}\right]\langle O^{--}(z_1,\bar{z}_1)O^{++}(z_2,\bar{z}_2)O_{\text{bos}}(z_3,\bar{z}_3)\bar{O}_{\text{bos}}(z_4,\bar{z}_4)\rangle ,
\end{split}
\end{equation}
where we used the action of the stress energy tensor on primaries in \eqref{eq:TTOPE}. Using the definition of the cross ratios \eqref{eq:crossratio} it is straightforward to write the derivative as
\begin{equation}
    \frac{\partial}{\partial z_i}=\frac{\partial z}{\partial z_i}\frac{\partial}{\partial z}+\frac{\partial \bar{z}}{\partial z_i}\frac{\partial}{\partial \bar{z}} \, ,
\end{equation}
such that in the $(z,\bar{z})$ coordinates the WI is realized as
\begin{equation}
\label{eq:WI3}
\begin{split}
\mathcal{G}^{(1,0,1)}(z,\bar{z})=\left[  (1-z)^2 \pd ( z \pd) + 1   \right] {\cal G} (z, \bar z) ,
\end{split}
\end{equation}
where $\mathcal{G}^{(1,0,1)}(z,\bar{z})$ and ${\cal G} (z, \bar z)$ are defined to be in the full LLLL correlators. Diagrammatically, we have
\begin{equation}
    \begin{tikzpicture}[scale=0.6, baseline={([yshift=-.5ex]current bounding box.center)},vertex/.style={anchor=base,
         circle}]]
    \draw (1,1) node[above]{$\phantom{1} \hspace{10mm} L_{-1} O^{--}$} --(0,0) -- (-1,-1) node[below]{$\bar{O}_B \hspace{2mm} \phantom{1}$};
    \draw (1,-1) node[below]{$\phantom{4} \hspace{10mm} L_{+1} O^{++}$}  --(0,0) -- (-1,+1) node[above]{$O_B \hspace{2mm} \phantom{1}$};
    \draw [thick] (0,0) circle [radius=1.414];
    \draw [fill, white] (0,0) circle [radius=0.6]; 
    \draw [pattern= horizontal lines, thick] (0,0) circle [radius=0.6]; 
    \end{tikzpicture} %
    = %
    |1-z|^4 \, \left[  (1-z)^2 \pd ( z \pd) + 1    \right] \left(    |1-z|^4 %
     \begin{tikzpicture}[scale=0.6, baseline={([yshift=-.5ex]current bounding box.center)},vertex/.style={anchor=base,
         circle}]]
    \draw (1,1) node[above]{$\phantom{1} \hspace{7mm} O^{--}$} --(0,0) -- (-1,-1) node[below]{$\bar{O}_B \hspace{2mm} \phantom{1}$};
    \draw (1,-1) node[below]{$\phantom{4} \hspace{7mm} O^{++}$}  --(0,0) -- (-1,+1) node[above]{$O_B \hspace{2mm} \phantom{1}$};
    \draw [thick] (0,0) circle [radius=1.414];
    \draw [fill, white] (0,0) circle [radius=0.6]; 
    \draw [pattern= horizontal lines, thick] (0,0) circle [radius=0.6]; 
    \end{tikzpicture}
    \right) .
\end{equation}

It can be checked from the explicit expressions \eqref{LLLL1} and \eqref{LLLL2} and by using the formula \eqref{eq:derDz} and the properties of the $D$-function that the WI works also in the $s$-channel and thus we can write 
\begin{equation}
\label{eq:WI4}
\begin{split}
\mathcal{G}^{(1,0,1)}_s(z,\bar{z})=\left[  (1-z)^2 \pd ( z \pd) + 1    \right] {\cal G}_s (z, \bar z) .
\end{split}
\end{equation}
Diagrammatically it is represented by
\begin{equation}
    \begin{tikzpicture}[scale=0.6, baseline={([yshift=-.5ex]current bounding box.center)},vertex/.style={anchor=base,
            circle}]] 
        \draw  (1,-1) node[below]{$\phantom{1} \hspace{10mm}   L_{+1} O^{++}$} -- (+0.6,0) -- (-0.6,0) -- (+0.6,0)  -- (1,1) node[above]{$\phantom{1} \hspace{10mm}   L_{-1} O^{--}$} ;
        \draw (-1,+1) node[above]{$ O_B \hspace{2mm} \phantom{1}$}   -- (-0.6,0) -- (+0.6,0) -- (-0.6,0) --   (-1,-1) node[below]{$\bar{O}_B \hspace{2mm} \phantom{1}$};
        \draw [thick] (0,0) circle [radius=1.414];
        \end{tikzpicture} %
     = %
    |1-z|^4 \, \left[  (1-z)^2 \pd ( z \pd) + 1    \right] \left(    |1-z|^4 %
        \begin{tikzpicture}[scale=0.6, baseline={([yshift=-.5ex]current bounding box.center)},vertex/.style={anchor=base,
         circle}]] 
        \draw  (1,-1) node[below]{$\phantom{1} \hspace{7mm} O^{++}$} -- (+0.6,0) -- (-0.6,0) -- (+0.6,0)  -- (1,1) node[above]{$\phantom{1} \hspace{7mm} O^{--}$} ;
        \draw (-1,+1) node[above]{$ O_B \hspace{2mm} \phantom{1}$}   -- (-0.6,0) -- (+0.6,0) -- (-0.6,0) --   (-1,-1) node[below]{$\bar{O}_B \hspace{2mm} \phantom{1}$};
        \draw [thick] (0,0) circle [radius=1.414];
        \end{tikzpicture} %
    \right)  .
\end{equation}
Similarly we can write the WI for spectrally-flowed state \eqref{q2} for the case $m=1$, i.e.
\begin{equation}
\label{eq:WI5}
\begin{split}
&\langle\left(J^+_{0}O^{--}(z_1,\bar{z}_1)\right)\left(J_{0}^-O^{++}(z_2,\bar{z}_2)\right)O_{\text{bos}}(z_3,\bar{z}_3)\bar{O}_{\text{bos}}(z_4,\bar{z}_4)\rangle .
\end{split}
\end{equation}
This can be done in the same way of the previous case by two insertion of the currents and by using the OPE relations \eqref{eq:JJOPE} to get the result
\begin{equation}
\label{eq:WI6}
\begin{split}
\mathcal{G}^{(1,1,0)}(z,\bar{z})={\cal G} (z, \bar z) ,
\end{split}
\end{equation}
i.e., diagrammatically, 
\begin{equation}
    \begin{tikzpicture}[scale=0.6, baseline={([yshift=-.5ex]current bounding box.center)},vertex/.style={anchor=base,
         circle}]]
    \draw (1,1) node[above]{$\phantom{1} \hspace{10mm} J_{0}^+ O^{--}$} --(0,0) -- (-1,-1) node[below]{$\bar{O}_B \hspace{2mm} \phantom{1}$};
    \draw (1,-1) node[below]{$\phantom{4} \hspace{10mm} J_0^- O^{++}$}  --(0,0) -- (-1,+1) node[above]{$O_B \hspace{2mm} \phantom{1}$};
    \draw [thick] (0,0) circle [radius=1.414];
    \draw [fill, white] (0,0) circle [radius=0.6]; 
    \draw [pattern= horizontal lines, thick] (0,0) circle [radius=0.6]; 
    \end{tikzpicture} %
    = %
     \begin{tikzpicture}[scale=0.6, baseline={([yshift=-.5ex]current bounding box.center)},vertex/.style={anchor=base,
         circle}]]
    \draw (1,1) node[above]{$\phantom{1} \hspace{7mm} O^{--}$} --(0,0) -- (-1,-1) node[below]{$\bar{O}_B \hspace{2mm} \phantom{1}$};
    \draw (1,-1) node[below]{$\phantom{4} \hspace{7mm} O^{++}$}  --(0,0) -- (-1,+1) node[above]{$O_B \hspace{2mm} \phantom{1}$};
    \draw [thick] (0,0) circle [radius=1.414];
    \draw [fill, white] (0,0) circle [radius=0.6]; 
    \draw [pattern= horizontal lines, thick] (0,0) circle [radius=0.6]; 
    \end{tikzpicture} .
\end{equation}
It is straightforward to check, using the explicit expression in \eqref{LLLL3} and \eqref{LLLL1}, that also in this case the WI {works also in the $s$-channel} and we have
\begin{equation}
\label{eq:WI7}
\begin{split}
\mathcal{G}^{(1,1,0)}_s(z,\bar{z})={\cal G}_s (z, \bar z) .
\end{split}
\end{equation}
These computations then furnish a highly non-trivial check to our results.

The last WI we want to recall, that will be useful in the next section, is the one coming from supersymmetry. This has been found in \cite{Bombini:2017sge} and it reads 
\begin{equation}
\label{eq:WI8}
\begin{split}
{\cal G}(z, \bar z)=|1-z|^4 \, \partial\bar{\partial}\left[\frac{\mathcal{G}^{\text{fer}}(z,\bar{z})}{|1-z|^2}\right] ,
\end{split}
\end{equation}
with $\mathcal{G}^{\text{fer}}(z,\bar{z})$ defined as
\begin{equation}
\begin{split}
    \langle O^{--}(z_1,\bar{z}_1)O^{++}(z_2,\bar{z}_2)O^{++}(z_3,\bar{z}_3)O^{--}(z_4,\bar{z}_4)\rangle=\frac{1}{|z_{12}|^2 |z_{34}|^2} \, \mathcal{G}^{\text{fer}}(z,\bar{z}) ,
\end{split}
\end{equation}
and the explicit expression has been found in eq.~(3.10) of \cite{Giusto:2018ovt} to be
\begin{equation}\label{eq:GFerFull}
    {\cal G}^{\text{fer}}(z,\bar{z}) = \left(1-\frac{1}{N} \right)(1+|1-z|^2)  + \frac{2}{\pi\, N} \,  |z|^2 |1-z|^2 (\hat{D}_{1122}+\hat{D}_{1212}+\hat{D}_{2112}) ,
\end{equation}
while the $s-$channel piece is the one reported in eq.~\eqref{eq:FERMres}. Again, diagrammatically it is
\begin{equation}\label{eq:SUSYWIWD}
    \begin{tikzpicture}[scale=0.6, baseline={([yshift=-.5ex]current bounding box.center)},vertex/.style={anchor=base,
         circle}]]
    \draw (1,1) node[above]{$\phantom{1} \hspace{7mm}  O^{--}$} --(0,0) -- (-1,-1) node[below]{$\bar{O}_B \hspace{2mm} \phantom{1}$};
    \draw (1,-1) node[below]{$\phantom{4} \hspace{7mm}  O^{++}$}  --(0,0) -- (-1,+1) node[above]{$O_B \hspace{2mm} \phantom{1}$};
    \draw [thick] (0,0) circle [radius=1.414];
    \draw [fill, white] (0,0) circle [radius=0.6]; 
    \draw [pattern= horizontal lines, thick] (0,0) circle [radius=0.6]; 
    \end{tikzpicture} %
    = %
    \partial \bar{\partial} \left(    %
     \begin{tikzpicture}[scale=0.6, baseline={([yshift=-.5ex]current bounding box.center)},vertex/.style={anchor=base,
         circle}]]
    \draw (1,1) node[above]{$\phantom{1} \hspace{7mm} O^{--}$} --(0,0) -- (-1,-1) node[below]{$O^{++} \hspace{7mm} \phantom{1}$};
    \draw (1,-1) node[below]{$\phantom{4} \hspace{7mm} O^{++}$}  --(0,0) -- (-1,+1) node[above]{$O^{--} \hspace{7mm} \phantom{1}$};
    \draw [thick] (0,0) circle [radius=1.414];
    \draw [fill, white] (0,0) circle [radius=0.6]; 
    \draw [pattern= horizontal lines, thick] (0,0) circle [radius=0.6]; 
    \end{tikzpicture}
    \right) .
\end{equation}
Also in this case the WI are {valid in the $s-$channel} and then we have 
\begin{equation}
\label{eq:WI8}
\begin{split}
{\cal G}_s(z, \bar z)=|1-z|^4\partial\bar{\partial}\left[\frac{\mathcal{G}_s^{\text{fer}}(z,\bar{z})}{|1-z|^2}\right] . 
\end{split}
\end{equation}

\subsection{Reconstructing the full LLLL correlators using the Ward identities}
So far we have used the validity of Ward identities in the $s$-channel to check our computations. Now we will instead use it to construct the full LLLL four-point functions involving the $\frac{1}{8}$-BPS operators. It is indeed easy to see that, via eq.~\eqref{eq:SUSYWIWD}, using eq.~\eqref{eq:GFerFull}, we have 
\begin{equation}
    {\cal G} (z, \bar z) =  1+  \frac{1}{\pi N}\left[  \left( 4(1  + |z|^2) \hat D_{3311} -2\pi \right) +  |1-z|^4  \hat D_{2222} \right] .
\end{equation}
From here, using eq.~\eqref{eq:WI6}, we can build 
\begin{equation}\label{eq:full110}
    {\cal G}^{(1,1,0)} (z, \bar z) =  1+  \frac{1}{\pi N}\left[  \left( 4(1  + |z|^2) \hat D_{3311} -2\pi \right) +  |1-z|^4  \hat D_{2222} \right] ,
\end{equation}
and, using eq.~\eqref{eq:WI3},
\begin{equation}
\begin{split}\label{eq:full101}
  {\cal G}^{(1,0,1)} (z, \bar z) =  1 + \frac{1}{\pi N} & \Big[  \left(4(1  + |z|^2) \hat D_{3311} -2\pi \right)  + 5   |1-z|^4  \hat D_{2222}   \\
  & \quad + 2 |1-z|^4   \frac{(\bar z \bar \pd^2 + \bar \pd)}{z} \left( |z|^2 \hat D_{2222} \right)  - 8  |1-z|^4 \hat{D}_{3322}      \Big] .
  \end{split}
\end{equation} 
These last two equations constitute the first example of all-light AdS$_3$ four-point functions at strong coupling involving $\frac{1}{8}$-BPS operators.

\section{Discussion}
\label{sec:disc}

In this paper we have holographically computed the four-point functions (\ref{eq:HHLL1}, \ref{eq:4pf110}, \ref{eq:4pfq1}) in the HHLL limit from a large family of $\frac{1}{8}$-BPS states whose dual supergravity solution is explicitly known. These four-point functions involve two heavy states \eqref{eq:3ch} and two Bosonic light states \eqref{eq:light}. In order to perform these computations we use a standard holographic technique: we identified the supergravity field  dual to the light state, and we consider it as a perturbation around the type IIB supergravity solution dual to the heavy state, that can be regarded as acting as a source for the equation of motion of the perturbation \cite{Galliani:2016cai, Galliani:2017jlg, Bombini:2017sge} (as in fig.~\ref{fig:HHLLWD}). This was possible to be achieved in the D1D5 CFT, due to the precise dictionary that was established between heavy states and type IIB solutions \cite{Bena:2016ypk, Bena:2017xbt, Bakhshaei:2018vux, Ceplak:2018pws}. We were also able to perform these computations without using the Witten diagram technology, that is still not properly defined in AdS$_3$  \cite{Heemskerk:2009pn, Giusto:2018ovt}, since, up to now, only the cubic coupling are known \cite{Arutyunov:2000by}. 

With respect to the analysis put forward in \cite{Bombini:2017sge}, we have computed the HHLL correlators on three-charge microstates of a black hole with non-degenerate horizon, in the sense that the ensemble of such states is described by a black hole with a finite horizon in the classical limit of supergravity \cite{Bena:2016ypk, Bena:2017xbt, Ceplak:2018pws}; our analysis was performed in the small-$b$ limit. In the regime under scrutiny, the correlators show a behaviour that is compatible with unitarity, as expected for pure states. It would then be very interesting to extend the results of \cite{Bombini:2017sge} in the geometries analyzed here, at least in the infinite throat limit, i.e. $a \ll b$, to see if the general mechanism for information conservation described in \cite{Bombini:2017sge} is extended to an ensemble dual to a regular black hole.

Recently, a conjecture on how to extract LLLL four-point functions from the HHLL one was put forward in \cite{Giusto:2018ovt}, where the authors suggest that, under a proper ``lightening'' limit, the HHLL correlator reduces to a single-channel contribution of the LLLL, let us say, the $s$-channel one. The reason  is quite simple: in the HHLL case, no single-trace operators are exchanged in the cross channels where one heavy and one light operator fuses, i.e., in this nomenclature, the $t$- and the $u$-channels; this implies that the HHLL correlator contains only contributions from the channel where the two heavy operators fuses. Crucial to this lightening limit was the fact that the theory contains two sectors, a Ramond and a Neveu-Schwarz sector, that are connected by a spectral flow.  We have shown in sec.~\ref{sec:LLLL} how spectral-flowing the one-parameter family of heavy states from the R sector to the NS sector and then taking an appropriate limit on the free parameter, it was possible to extract the aforementioned $s$-channel contribution of the LLLL correlator. 

This conjecture is further supported by looking at the correlator in Mellin space (see app.~\ref{sec:AppD}); in fact, from the Mellin transform of the four-point functions we can read all the single-trace operators that are exchanged simply by looking at their poles \cite{Penedones:2010ue}. The location of the poles describes the twist of the single trace, while the residue at the pole is related to the three-point function at the vertexes. It is easy to see that the $s$-channel contribution deduced from the HHLL correlator \eqref{eq:FERMres} reads in Mellin space 
\begin{equation}
    M_s^{\rm fer} (s,t,u)=-2\left(1-\frac{t}{2}\right)^2\frac{1}{s} \,,
\end{equation}
showing then only the pole in $s$. Since this correlator is of the form $\bra O (z_1) \bar O (z_2) \bar O (z_3)  O (z_4) \ket $, one may expect that also a $u$-channel pole should arise, due to the evident symmetry under $z_1 \leftrightarrow z_3$. Indeed, the fully reconstructed one \eqref{eq:GFerFull} in Mellin space reads
\begin{equation}
    M^{\rm fer}(s,t,u)= \left(1-\frac{t}{2}\right)-2\left(1-\frac{t}{2}\right)^2\left(\frac{1}{s}+\frac{1}{u}\right),
\end{equation}
showing clearly the $u$-channel pole as well as all the correct symmetries, i.e. $s\leftrightarrow u$.  The only poles arising are then $s=0$ and $u=0$. As already pointed out, from here we read that the exchanged single-traces have zero twist and correspond therefore to conserved currents \cite{Galliani:2017jlg}.  

We have then checked the results we have found by using Ward identities that relate the correlators. This was possible since in the R sector the heavy states are generated by acting with certain generators of the global part of the  superconformal Kac-Moody algebra. This constitutes a non-trivial check on our computations, but, maybe more interestingly, it furnishes a way to reconstruct the full LLLL correlators (\ref{eq:full110}, \ref{eq:full101}), using the result of \cite{Giusto:2018ovt}, eq.~\eqref{eq:GFerFull}, as a seed.  

We can then compute the Mellin transform of both the $s$-channel result \eqref{LLLL3} for the $\bra (J_0^+ O^{--})(J_0^- O^{++}) O_B \bar O_B   \ket$ correlator, that reads
\begin{equation}
    M_s^{(1,1,0)} (s,t,u)=+ (s-2)-\frac{(t-3)^2}{s}-\frac{(u-3)^2}{s} \,,
\end{equation}
as well as the full LLLL result \eqref{eq:full110}; 
\begin{equation}
    M^{(1,1,0)} (s,t,u)= -(s-2)-\frac{(t-3)^2}{s}-\frac{(u-3)^2}{s} \,.
\end{equation}
These result manifestly have the right symmetry, that is $z_3 \leftrightarrow z_4$, i.e. $t \leftrightarrow u$. Also, we see that, again, only twist-zero single-trace operators are exchanged, namely the identity and the affine currents. From the latter we also learn that there is no exchange of single-trace operators in the $t$- and $u$-channels, meaning that the only difference among the $s$-channel part and the full LLLL is a term containing no poles, that can be interpreted as a contact term \cite{Heemskerk:2009pn}. 

All these results were found without using the Witten diagram technology; on the contrary, we hope that this procedure may be helpful to shed light on the structure of Witten diagrams in AdS$_3$. In fact, we may deduce out from the correlators we have computed the values of the Witten diagrams by looking backwards, namely reading the single diagram contribution to the amplitude, effectively reconstructing (some of) the three- and four- point vertexes.

\vspace{7mm}
 \noindent {\large \textbf{Acknowledgements} }

 \vspace{5mm} 

We would like to thank R.~Russo for valuable discussions, D.~Turton for useful observations and for having brought some references to our attention, and especially S.~Giusto for useful comments on the draft. AB is supported by the ANR grant Black-dS-String ANR-16-CE31-0004-01. AG is supported  by the John Templeton Foundation grant 61169. AB wishes to thank the Institut de Physique Th\'eorique, Universit\'e Paris Saclay, CEA, for the hospitality while this project was realized.


\appendix

\section{The D1D5 CFT}
\label{sec:AppA}
At the orbifold the D1D5 CFT has a realization in terms of free fields. We follow the conventions of~\cite{Avery:2010qw, Galliani:2016cai, Ceplak:2018pws}. In particular we have the Operator Product Expansion (OPE) between the elementary fields
\begin{equation}
  \label{eq:elopes}
  \psi^{\alpha\dot{A}}_{(r)}(z)\,\psi^{\beta\dot{B}}_{(s)}(w) \sim - \frac{\epsilon^{\alpha\beta}\,\epsilon^{\dot{A}\dot{B}}\,\delta_{r,s}}{z-w}  
~,~~~
\partial X^{A\dot{A}}_{(r)}(z)\,\partial X^{B\dot{B}}_{(s)}(w) \sim \frac{\epsilon^{AB}\epsilon^{\dot{A}\dot{B}}\,\delta_{r,s}}{(z-w)^2}  \,,
\end{equation}
where the $SU(2)$ indices are raised and lowered by using the $\epsilon$ tensor with the convention $\epsilon_{12} = - \epsilon_{21} = \epsilon^{21} = -\epsilon^{12} = +1$, for instance
\begin{equation}
\partial X_{A\dot{A}} = \epsilon_{AB} \epsilon_{\dot{A}\dot{B}}\,\partial X^{B\dot{B}},\qquad \partial X^{A\dot{A}} = \epsilon^{AB} \epsilon^{\dot{A}\dot{B}}\,\partial X_{B\dot{B}} \, ,
\end{equation}
and similarly for the antiholomorphic fields. The currents generating the algebra of the symmetries of the theory have a free field realization in the orbifold theory as
 \begin{equation}
\begin{aligned}
\label{eq:OPE1}
T(z)&=\frac{1}{2}\sum_{r=1}^N\epsilon_{\dot{A}\dot{B}}\epsilon_{AB}\partial X_{(r)}^{A\dot{A}}\partial X_{(r)}^{B\dot{B}}+\frac{1}{2}\sum_{r=1}^N\epsilon_{\alpha\beta}\epsilon_{\dot{A}\dot{B}}\psi_{(r)}^{\alpha\dot{A}}\partial\psi_{(r)}^{\beta\dot{B}} \,, \\
G^{\alpha A}(z)&=\sum_{r=1}^N\psi_{(r)}^{\alpha\dot{A}}\partial X^{\dot{B}A}_{(r)}\epsilon_{\dot{A}\dot{B}} \,,\\
J^{a}(z)&=\frac{1}{4}\sum_{r=1}^N\epsilon_{\dot{A}\dot{B}}\psi_{(r)}^{\alpha\dot{A}}\epsilon_{\alpha\beta}(\sigma^{*a})^{\beta}{}_{\gamma}\psi_{(r)}^{\gamma\dot{B}} \,.
\end{aligned}
\end{equation}
and they satisfies the following OPEs \cite{Avery:2010qw}
\begin{subequations}\label{eq:OPE2}
\begin{align}
T(z)T(w)&\sim \frac{c/2}{(z-w)^4}+\frac{2T(w)}{(z-w)^2}+\frac{\partial T(w)}{(z-w)}  \label{eq:TTOPE} \,,\\
G^{\alpha A}(z)G^{\beta B}(w)&\sim -\frac{c}{3}\frac{\epsilon^{\alpha\beta}\epsilon^{AB}}{(z-w)^3}+\epsilon^{AB}\epsilon^{\beta\gamma}(\sigma^{*a})^{\alpha}_{\gamma}\left[\frac{2J^a(w)}{(z-w)^2}+\frac{\partial J^a(w)}{(z-w)}\right]-\epsilon^{\alpha\beta}\epsilon^{AB}\frac{T(w)}{(z-w)} \,, \label{eq:GGOPE}\\
J^a(z)J^b(w)&\sim\frac{c}{12}\frac{\delta^{ab}}{(z-w)^2}+i\epsilon^{ab}{}_c\frac{J^c(w)}{(z-w)} \,,  \label{eq:JJOPE}\\
T(z)J^a(w)&\sim\frac{J^a(w)}{(z-w)^2}+\frac{\partial J^a(w)}{(z-w)}\\
T(z)G^{\alpha A}(w)&\sim\frac{3}{2}\frac{G^{\alpha A}(w)}{(z-w)^2}+\frac{\partial G^{\alpha A}(w)}{(z-w)} \,, \\
J^a(z)G^{\alpha A}(w)&\sim\frac{1}{2}(\sigma^{*a})^{\alpha}_{\beta}\frac{G^{\beta A}(w)}{(z-w)} \,. 
\end{align}
\end{subequations}
As standard in two-dimensional CFT we can expand the currents in modes
\begin{equation}
\label{eq:exp}
\begin{aligned}
T(z)&=\sum_nL_nz^{-n-2},\qquad L_n=\oint\frac{dz}{2\pi i} \, T(z)z^{n+1} \,, \\
J^a(z)&=\sum_nJ^a_nz^{-n-1},\qquad J^a_n=\oint\frac{dz}{2\pi i} \, J^a(z)z^{n}  \,,  \\
G^{\alpha A}(z)&=\sum_nG^{\alpha A}_nz^{-n-\frac{3}{2}},\quad G^{\alpha A}_n=\oint\frac{dz}{2\pi i} \, G^{\alpha A}(z)z^{n+\half} \,.
\end{aligned}
\end{equation}
Using this expansion and the OPEs we can write down the affine algebra \cite{Ceplak:2018pws}
\begin{subequations}
\label{eq:affinealg}
\begin{align}
\left[ L_n, L_m \right] &= (n-m) L_{n+m}\,, \quad  \left[ J^a_0, J^b_0 \right] = i\epsilon^{abc} J^c_{0} \,, \quad [L_n, J_0^a] =0 \,,\\
\left\{G^{\alpha A}_r, G^{\beta B}_s\right\} &= \ep^{\alpha \beta} \ep^{AB} L_{r+s} + (r-s) \ep^{AB} (\sigma^{a T})^\alpha{}_\gamma \ep^{\gamma \beta} J^a_{r+s} \,, \\
\left[ J^a_0, G^{\alpha A}_s \right] &= \frac{1}{2} \, G^{\beta A}_{s} (\sigma^{a})_{\beta}{}^{\alpha} \,, \quad 
\left[ L_m, G^{\alpha A}_s \right] = \left(\frac{m}{2}-s\right)G^{\alpha A}_{m+s}  \,.
\end{align}
\end{subequations}
We reported here only the global part, i.e. $n,m=-1,0,1$ and $r,s= \pm \half$, that is the only one that we used in the paper. Also, our conventions for the Levi-Civita tensor are $\ep_{12}=\ep_{+-}=\ep^{-+}=\ep^{21} =+1$. 
The action of the current algebra on primary operators reads
\begin{equation}
    \begin{split}
        T(z)O_p(w)&=\frac{h_p}{(z-w)^2} \, O_p(w)+\frac{1}{z-w} \, \partial O_p(w)+ [\text{reg}] \,, \\
        J^a(z) O^{\alpha\dot{\alpha}}_p(w)&= \half \, \frac{ (\sigma^{*a})^\alpha{}_\beta }{z-w}  \,  O^{\beta \dot{\alpha}}_p(w)  + [\text{reg}] \,.
    \end{split}
\end{equation}

Notice that the OPEs and the action of the current algebra on primary fields are defined to be the same even at the supergravity point \cite{Avery:2010qw, Galliani:2016cai, Ceplak:2018pws}, not only at the free-orbifold point.


\section{Bulk integrals}
\label{sec:AppB}

In this appendix we describe of to compute Bulk integrals like the ones appearing in eq.~\eqref{eq:BulkIntC110}. In order to do so, we need to introduce the Bulk-to-Boundary propagator of conformal dimension $\Delta$ as
\begin{equation}
    B_0 ({\bf r}' | \tau_E \,, \sigma) \equiv K_\Delta ({\bf r}' | \tau_E \,, \sigma) = \left[ \half \, \frac{a_0}{\sqrt{r^2+a_0^2} \, \cos (\tau_E' -\tau_E) - r \, \sin (\sigma' - \sigma)}   \right]^\Delta \,;
\end{equation}
with this notation, we read $B_0 ({\bf r}) \equiv  B_0 ({\bf r}' | 0,0)$. We also have that 
\begin{equation}
    B_{\pm} \equiv \frac{a_0}{\sqrt{r^2+a_0^2}} \, e^{\pm \tau_E} \,,
\end{equation}
are the bulk-to-boundary propagators with $\Delta = 1$ evaluated at the points $z = \infty$ and $z = 0$, i.e. 
\begin{subequations}
\begin{align}
& B_0 ({\bf r}' | \tau_E , \sigma) = |z|^2 K_2 ({\bf w} | z, \bar z) ,\\
& B_+ ({\bf r}') = \lim_{z_2\rightarrow \infty} |z_2|^2 K_1 ({\bf w} | z_2, \bar z_2) \equiv K_1 ( {\bf w} |\infty) = w_0\, , \\
& B_-({\bf r}') = K_1 ({\bf w} | 0) \, ,
\end{align}
\end{subequations}
where ${\bf w} = \{ w_0, w, \bar w \}$ are the AdS$_3$ Poincar\'e patch coordinates \eqref{eq:AdSdPc}.  We will also use that
\begin{equation}
    - 2\, a_0^2 \frac{r}{r^2+a_0^2} \, \pd_r B_0 = (B_+ \pd_\mu B_ + B_- \pd_\mu B_+) \pd^\mu B_0 \,.
\end{equation}
The bulk integrals that appear in the text are then
\begin{subequations}
\begin{align}
    I_1&\equiv \int d^3 {\bf r}'_e \sqrt{\bar g} \,B_0({\bf r}'_e|\tau_E, \sigma)\,\partial'^{\mu}B_0({\bf r}'_e|0,0)\,B_-({\bf r}'_e)\,\partial'_\mu B_+({\bf r}'_e)\,,\\
    I_2&\equiv \int d^3 {\bf r}'_e \sqrt{\bar g} \,B_0({\bf r}'_e|\tau_E, \sigma)\,\partial'^{\mu}B_0({\bf r}'_e|0,0)\,B_+({\bf r}'_e)\,\partial'_\mu B_-({\bf r}'_e)\,,\\
    I_3&\equiv \int d^3 {\bf r}'_e \sqrt{\bar g} \,B_0({\bf r}'_e|\tau_E, \sigma)\, \partial^2_{\tau'_E}B_0({\bf r}'_e|0,0)\,\frac{a_0^4}{(r'^2+a_0^2)^2}\, , \\
    I^{(p)} & \equiv \int d^3 {\bf r}'_e \sqrt{\bar g} \,B_0({\bf r}'_e|\tau_E, \sigma)\, B_0({\bf r}'_e|0,0)\,\left(\frac{a_0^2}{r'^2+a_0^2} \right)^p \,, 
\end{align}
\end{subequations}
These integrals can be written in terms of the same $D$-functions $D_{\Delta_1 \Delta_2 \Delta_3 \Delta_4}$, defined in eq.~\eqref{eq:Dintg}, that usually appear in the computations of Witten diagrams.

The first integral can be computed as in \cite{Galliani:2017jlg} by writing it in Poincar\'e coordinates as
\begin{equation}
\begin{aligned}
    |z|^{-2} I_1 =&\, \int \! d^3\mathbf{w}\,w_0^{-1} \left(\frac{w_0}{w_0^2+|w-z|^2}\right)^2 \left[ \frac{2 w_0}{(w_0^2 + |w-1|^2)^2}- \frac{4 w_0^3}{(w_0^2 + |w-1|^2)^3}\right] \frac{w_0}{w_0^2+|z|^2}\\
    =&\,2 \hat D_{1122}-4 \hat D_{1232}\,.
\end{aligned}
\end{equation}
Similarly we get
\begin{equation}
|z|^{-2} I_2 = 2 \hat D_{2222}-2 \hat D_{1122}+4 \hat D_{1232}\,. 
\end{equation}
The computation of $I_3$ instead is done noticing that we can integrate by parts exchanging $\partial_{\tau'_E}^2$ with $- \partial_{\tau_E}^2$, since everything depends only on the difference $\tau_E' - \tau_E$:
\begin{equation}\label{eq: identity}
\begin{aligned}
I_3 &=  \partial_{\tau_E} \frac{I_1-I_2}{2} = (z \partial + \bar z \bar \partial)\left(|z|^2(2 \hat D_{1122}-4 \hat D_{1232}- \hat D_{2222})\right)\\
&= \frac{2 |z|^2}{|1-z|^4}\left(2 (1+|z|^2) \hat D_{3311}-\pi \right)\,,
\end{aligned}
\end{equation}
where the last identity follows from a computation that uses properties of the $\hat D$-functions reported in app.~\ref{sec:AppC}.

The last integral is similarly computed as
\begin{equation}
    I^{(p)}  = |z|^2 \hat{D}_{pp22} \,.
\end{equation}
Now, using again  that we can integrate by parts exchanging $\partial_{\tau'_E}^2$ with $- \partial_{\tau_E}^2$, since everything depends only on the difference $\tau_E' - \tau_E$, we can compute 
\begin{equation}
    \tilde I^{(p)}  \equiv \int d^3 {\bf r}'_e \sqrt{\bar g} \,B_0({\bf r}'_e|\tau_E,y)\,\pd_i \pd_j B_0({\bf r}'_e|0,0)\,\left(\frac{a_0^2}{r'^2+a_0^2} \right)^p \,,
\end{equation}
with $i$,$j$ that can be either $\tau$ or $\sigma$, giving
\begin{equation}
    \tilde I^{(p)} =  \pd_i \pd_j \left(  |z|^2 \hat{D}_{pp22}  \right) .
\end{equation}


\section{D-function technology}
\label{sec:AppC}
The $D$-functions are defined as \cite{Dolan:2000ut, Arutyunov:2002fh}
\begin{equation}
  \label{eq:Dintg}
  D_{\Delta_1 \Delta_2 \Delta_3 \Delta_4}(z_i) = \int d^{d+1}w \, \sqrt{\bar g}\, \prod_{i=1}^4 K_{\Delta_i}(w,\vec{z}_i)\;,
\end{equation}
where the AdS$_{d+1}$ metric in the Euclidean Poincar\'e coordinates is\footnote{The change of coordinates that brings global AdS$_3$ into this set is 
\begin{equation*}
    w_0 = \frac{a_0}{\sqrt{r^2+a_0^2}} \, e^{i \tau} \,, \quad w = \frac{r}{\sqrt{r^2+a_0^2}} \, e^{i (\tau + \sigma)} \,, \quad \bar w = \frac{r}{\sqrt{r^2+a_0^2}} \, e^{i (\tau - \sigma)} \,.
\end{equation*}}
\begin{equation}
  \label{eq:AdSdPc}
  d{\bar s}^2 = \frac{1}{w_0^2} \left(dw_0^2 + \sum_{i=1}^d dw_i^2 \right)~.
\end{equation}
and the unnormalised boundary-to-bulk propagator for a scalar field propagating in Euclidean AdS$_{d+1}$ is
\begin{equation}
  \label{eq:b2bd}
  K_{\Delta}(w,\vec{z}) = 
\left[\frac{w_0}{w_0^2 + (\vec{w}-\vec{z})^2} \right]^{\Delta}=\frac{
1}{\Gamma(\Delta)} \int_0^\infty \!\! dt\, w_0^{\Delta}\, t^{\Delta-1} e^{-t(w_0^2+(\vec{w}-\vec{z})^2)}\;,
\end{equation}
where $\Delta$ is the conformal dimension of the dual operator
By using the representation of the propagator in terms of Schwinger parameters given in~\eqref{eq:b2bd} it is straightfoward to perform the integration over the interaction point $(w_0,\vec{w})$ and obtain
\begin{equation}
  \label{eq:Dintg2}
   D_{\Delta_1 \Delta_2 \Delta_3 \Delta_4}(z_i) = \Gamma\left(\frac{\hat{\Delta}-d}{2}\right)\int_0^\infty \prod_i\left[ dt_i \, \frac{t_i^{\Delta_i-1}}{\Gamma(\Delta_i)} \right] \frac{\pi^{d/2}}{2 T^{\frac{\hat\Delta}{2}}} e^{-\sum_{i,j=1}^4 |z_{ij}|^2 \frac{t_i t_j}{2 T}} \, ,
\end{equation}
with $T= \sum_i t_i$, $\hat\Delta=\sum_i \Delta_i$ and ${z}_{ij}={z}_i - {z}_j$, where $z_j$ are the standard complex coordinates. 

We also define the cross ratios
\begin{equation}
    \label{eq:crossratio}
    u=(1-z)(1-\bar{z})=\frac{z_{12}^2z_{34}^2}{z_{13}^2z_{24}^2},\quad  v=z\bar{z}=\frac{z_{14}^2z_{23}^2}{z_{13}^2z_{24}^2} \,.
\end{equation}
The $\hat{D}$-functions which appear in the bulk computation of the correlators are defined as
\begin{equation}
    \label{eq:Dhat}
    \hat{D}_{\Delta_1 \Delta_2 \Delta_3 \Delta_4}(z,\bar{z})=\lim_{z_2\to\infty}|z_2|^{2\Delta_2}D_{\Delta_1 \Delta_2 \Delta_3 \Delta_4}(z_1=0,z_2,z_3=1,z_4=z) .
\end{equation}
Once written in terms of Schwinger parameter, one can see that $D_{1111}$ is proportional to the massless box-integral in four dimensions with external massive state, the result can be written in term of logarithms and dilogarithms
\begin{equation}
  \label{eq:D1111}
  D_{1111}= 
\frac{ \pi}{2 |z_{13}|^2 |z_{24}|^2 (z-\bar{z})} \left[2{\rm Li}_2(z) - 2 {\rm Li}_2(\bar{z}) + \ln(z \bar{z}) \ln\frac{1-z}{1-\bar{z}}\right]~,
\end{equation}
The result in~\eqref{eq:D1111} is proportional to the Bloch-Wigner dilogarithm $D(z,\bar{z})$ \cite{Galliani:2017jlg, Bombini:2017sge}
\begin{equation}
  \label{eq:D1111bis}
   D_{1111}= 
\frac{2\pi i}{|z_{13}|^2 |z_{24}|^2 (z-\bar{z})} D(z,\bar{z})\, ,
\end{equation}
where
\begin{equation}
  \label{eq:BWD}
  \begin{aligned}
  D(z,\bar{z}) & =  {\rm Im} [{\rm Li}_2(z)] + {\rm Arg}[\ln(1-z)] \ln|z| 
 \\ & =  \frac{1}{2i} \left[{{\rm Li}_2(z)-{\rm Li}_2(\bar{z})+ \frac{1}{2} \ln(z\bar{z}) \ln\frac{1-z}{1-\bar{z}} }\right]\,.
  \end{aligned}
\end{equation}
Moreover we have the following useful identities
\begin{equation}
  \label{eq:BWiden}
  D(z,\bar{z}) = - D\left(\frac{1}{z},\frac{1}{\bar{z}}\right) =  
  - D\left({1-z},{1-\bar{z}}\right) ~,
\end{equation}
which implies
\begin{equation}
  \label{eq:BWiden2}
 D(z,\bar{z}) = D\left(1-\frac{1}{z},1-\frac{1}{\bar{z}}\right) =  
  D\left(\frac{1}{1-z},\frac{1}{1-\bar{z}}\right) =  
  - D\left(\frac{-z}{1-z},\frac{-\bar{z}}{1-\bar{z}}\right) ~.
\end{equation}
Other useful relations are \cite{Giusto:2018ovt}
\begin{subequations}
\begin{align}
\hat{D}_{\Delta_2 \Delta_1 \Delta_3 \Delta_4}\left(\frac{1}{z},\frac{1}{\bar{z}}\right)&=|z|^{2\Delta_4}\hat{D}_{\Delta_1 \Delta_2 \Delta_3 \Delta_4}(z,\bar{z}) \label{Drel1}, \\
\hat{D}_{\Delta_3 \Delta_2 \Delta_1 \Delta_4}\left(1-z,1-\bar{z}\right)&=\hat{D}_{\Delta_1 \Delta_2 \Delta_3 \Delta_4}(z,\bar{z}) , \label{Drel2}\\
\hat{D}_{\Delta_2 \Delta_1 \Delta_4 \Delta_3}\left(z,\bar{z}\right)&=|z|^{\Delta_1-\Delta_2-\Delta_3+\Delta_4}\hat{D}_{\Delta_1 \Delta_2 \Delta_3 \Delta_4}(z,\bar{z}) . \label{Drel3}
\end{align}
\end{subequations}
Using in order \eqref{Drel2}, \eqref{Drel3}, \eqref{Drel2} we get the useful relation
\begin{equation}\label{Drel4}
    \hat D_{\Delta_1 \Delta_2 \Delta_3 \Delta_4} (z, \bar z) = |1-z|^{\Delta_2 - \Delta_3 - \Delta_4 + \Delta_1} \hat D_{\Delta_4 \Delta_3 \Delta_2 \Delta_1} (z, \bar z) .
\end{equation}
In the literature, it is customary to introduce the  $\bar{D}$-functions, which depend only on the cross ratios \cite{Penedones:2010ue, Dolan:2000ut, Arutyunov:2002fh}; we will briefly review how they are defined in our notation, where they read
\begin{equation}
    \label{eq:Dbar}
    \bar{D}_{\Delta_1 \Delta_2 \Delta_3 \Delta_4}(z,\bar{z})=\frac{2\prod_{i=1}^4\Gamma(\Delta_i)}{\pi^{d/2}\Gamma\left(\frac{\hat{\Delta}-d}{2}\right)}\frac{|z_{13}|^{\hat{\Delta}-2\Delta_4}|z_{24}|^{2\Delta_2}}{|z_{14}|^{\hat{\Delta}-2\Delta_1-2\Delta_4}|z_{34}|^{\hat{\Delta}-2\Delta_3-2\Delta_4}}D_{\Delta_1 \Delta_2 \Delta_3 \Delta_4}(z_i) .
\end{equation}
The relation between $\hat{D}$ and $\bar{D}$-functions is
\begin{equation}
    \label{eq:relD1}
    \hat{D}_{\Delta_1 \Delta_2 \Delta_3 \Delta_4}(z,\bar{z})=\frac{\pi\Gamma\left(\frac{\hat{\Delta}-2}{2}\right)}{2\prod_{i=1}^4\Gamma(\Delta_i)}|z|^{\hat{\Delta}-2\Delta_1-2\Delta_4}|1-z|^{\hat{\Delta}-2\Delta_3-2\Delta_4}\bar{D}_{\Delta_1 \Delta_2 \Delta_3 \Delta_4}(z,\bar{z}) .
\end{equation}

In the correlators also appears the derivative of the $\hat{D}$-functions with respect $z$ or $\bar{z}$. In order to handle with $\hat{D}$-functions it is useful to write these contributions in terms of $\hat{D}$-functions without derivatives. In order to find an useful expression we rewrite the generic $\hat{D}$-functions in terms of $\bar{D}$-functions using \eqref{eq:relD1} and \eqref{eq:Dbar}, and expressing them as function of $u$ and $v$. Thus we have
\begin{equation}
    \label{eq:relDuv}
    \hat{D}_{\Delta_1 \Delta_2 \Delta_3 \Delta_4}(u,v)=\frac{\pi\Gamma\left(\frac{\hat{\Delta}-2}{2}\right)}{2\prod_{i=1}^4\Gamma(\Delta_i)}v^{\frac{\hat{\Delta}-2\Delta_1-2\Delta_4}{2}}u^{\frac{\hat{\Delta}-2\Delta_3-2\Delta_4}{2}}\bar{D}_{\Delta_1 \Delta_2 \Delta_3 \Delta_4}(u,v) .
\end{equation}
The object to be computed is therefore
\begin{align}
    \frac{\partial}{\partial z}\hat{D}_{\Delta_1 \Delta_2 \Delta_3 \Delta_4}(z,\bar{z})=\left(\frac{\partial v}{\partial z}\frac{\partial}{\partial v}+\frac{\partial u}{\partial z}\frac{\partial}{\partial u}\right)\hat{D}_{\Delta_1 \Delta_2 \Delta_3 \Delta_4}(u,v) .
\end{align}
Rewriting now the $\hat{D}_{\Delta_1 \Delta_2 \Delta_3 \Delta_4}(u,v)$ in terms of $\bar{D}_{\Delta_1 \Delta_2 \Delta_3 \Delta_4}(u,v)$ using \eqref{eq:relDuv} and $\frac{\partial v}{\partial z}=\bar{z}$, $\frac{\partial u}{\partial z}=-(1-\bar{z})$ we end up with an expression containing $\bar{D}_{\Delta_1 \Delta_2 \Delta_3 \Delta_4}(u,v)$-functions and its derivatives with respect to $u$ and $v$. We can use the result of \cite{Dolan:2000ut, Arutyunov:2002fh, Gary:2009ae}:
\begin{align}
    \frac{\partial}{\partial u}\bar{D}_{\Delta_1 \Delta_2 \Delta_3 \Delta_4}(u,v)&=-\bar{D}_{\Delta_1+1 \Delta_2+1 \Delta_3 \Delta_4}(u,v) , \\
    \frac{\partial}{\partial v}\bar{D}_{\Delta_1 \Delta_2 \Delta_3 \Delta_4}(u,v)&=-\bar{D}_{\Delta_1 \Delta_2+1 \Delta_3+1 \Delta_4}(u,v) .
\end{align}
Reconstructing now the $\hat{D}$-functions and coming back to $z,\bar{z}$ dependence we get
\begin{align}
\label{eq:derDz}
  \frac{\partial}{\partial z}\hat{D}_{\Delta_1 \Delta_2 \Delta_3 \Delta_4}(z,\bar{z})&=\left(\frac{\hat{\Delta}-2\Delta_1-2\Delta_4}{2z(1-z)}+\frac{\Delta_4-\Delta_2}{1-z}\right)\hat{D}_{\Delta_1 \Delta_2 \Delta_3 \Delta_4}(z,\bar{z})\nonumber\\
  &\quad +\frac{2\Delta_2\Delta_3}{(2-\hat{\Delta})z}\hat{D}_{\Delta_1 \Delta_2+1 \Delta_3+1 \Delta_4}(z,\bar{z})\\
  &\quad -\frac{2\Delta_1\Delta_2}{(2-\hat{\Delta})(1-z)}\hat{D}_{\Delta_1+1 \Delta_2+1 \Delta_3 \Delta_4}(z,\bar{z}) . \nonumber
\end{align}
With very similar procedure it's straightforward to find 
\begin{align}
\label{eq:derDzb}
  \frac{\partial}{\partial\bar{z}}\hat{D}_{\Delta_1 \Delta_2 \Delta_3 \Delta_4}(z,\bar{z})&=\left(\frac{\hat{\Delta}-2\Delta_1-2\Delta_4}{2\bar{z}(1-\bar{z})}+\frac{\Delta_4-\Delta_2}{1-\bar{z}}\right)\hat{D}_{\Delta_1 \Delta_2 \Delta_3 \Delta_4}(z,\bar{z})\nonumber\\
  &\quad +\frac{2\Delta_2\Delta_3}{(2-\hat{\Delta})\bar{z}}\hat{D}_{\Delta_1 \Delta_2+1 \Delta_3+1 \Delta_4}(z,\bar{z})\\
  &\quad -\frac{2\Delta_1\Delta_2}{(2-\hat{\Delta})(1-\bar{z})}\hat{D}_{\Delta_1+1 \Delta_2+1 \Delta_3 \Delta_4}(z,\bar{z}) . \nonumber
\end{align}
The above results are enough to find the generic term containing higher derivatives of $\hat{D}$-functions and to trade it as a sum of non-derivative $\hat{D}$-functions. 

Acting with $\pd_z$ on the relations \eqref{Drel3}, \eqref{Drel4}, and using them, can give us non-trivial relations among D-functions of different $\hat \Delta$, as
\begin{equation}
    \begin{split}
        \hat D_{2442} (z, \bar z) &= \frac{4}{9} \left( \hat D_{2332} (z, \bar z) + |z|^2 \hat D_{3333} (z, \bar z) \right) , \\
        \hat D_{4422} (z, \bar z) &= \frac{4}{9} \left( \hat D_{3322} (z, \bar z) + |1-z|^2 \hat D_{3333} (z, \bar z) \right) .
    \end{split}
\end{equation}


\section{Mellin representation}
\label{sec:AppD}
Following \cite{Penedones:2010ue}, the Mellin amplitudes for four-point function 
\begin{align}
\label{e0}
\mathcal{C}_{\Delta_1\Delta_2\Delta_3\Delta_4}\left(z_i\right)=\langle O_1(z_1) O_2 (z_2) O_3 (z_3) O_4(z_4)\rangle \,,
\end{align}
is defined as
\begin{align}
\label{e1}
\mathcal{C}_{\Delta_1\Delta_2\Delta_3\Delta_4}\left(z_i\right)= \prod_{i<j}^{4}\int_{c-i\infty}^{c+i\infty}\frac{ds_{ij}}{2\pi i}M\left(s_{ij}\right)\Gamma(s_{ij})(z_{ij}^2)^{-s_{ij}} , 
\end{align}
with the following constraint
\begin{align}
\label{ee1}
\sum_{j=1,j\neq i}^4s_{ij}=\Delta_i  \,.
\end{align}
The variables $s_{ij}$ can be defined introducing $4$ Lorentzian vectors $p_i$ satisfying 
\begin{subequations}
\begin{align}
\sum_{i=1}^4p_i &= 0\, , \quad\quad p_i^2=-\Delta_i \,,  \\
s_{ij}& = p_i\cdot p_j=\frac{1}{2}(\Delta_i+\Delta_j+(p_i+p_j)^2) \label{eq:sij}.
\end{align}
\end{subequations}
In order to compute the Mellin transform of D-functions let us recall the definition \eqref{eq:Dintg2}
\begin{equation}
   D_{\Delta_1 \Delta_2 \Delta_3 \Delta_4}(z_i) = \Gamma\left(\frac{\hat{\Delta}-d}{2}\right)\int_0^\infty \prod_i\left[ dt_i \,  \frac{t_i^{\Delta_i-1}}{\Gamma(\Delta_i)} \right] \frac{\pi^{d/2}}{2 T^{\frac{\hat\Delta}{2}}} \, e^{-\sum_{i,j=1}^4 z_{ij}^2 \frac{t_i t_j}{2 T}} \,,
\end{equation}
with $T= \sum_i t_i$, $\hat\Delta=\sum_i \Delta_i$. Performing the following change of variables
\begin{align}
t_i\longrightarrow T^{\frac{1}{2}} \, t_i \,, 
\end{align}
the D-function takes the form 
\begin{align}
\label{eq:e6}
D_{\Delta_1 \Delta_2 \Delta_3 \Delta_4}(z_i)&=\pi^{d/2}\Gamma\left(\frac{\hat{\Delta}-d}{2}\right)\int_0^\infty \prod_i\left[ dt_i \,  \frac{t_i^{\Delta_i-1}}{\Gamma(\Delta_i)} \right]  e^{-\sum_{i,j=1}^4 z_{ij}^2 t_i t_j} \, .
\end{align}
By rescaling the coordinates
\begin{subequations}
\begin{align}
t_1&=\frac{|z_{23}|}{|z_{12}||z_{13}|} \, \hat{t}_1,\quad t_2=\frac{|z_{13}|}{|z_{12}||z_{23}|} \, \hat{t}_2,\quad t_3=\frac{|z_{12}|}{|z_{13}||z_{23}|} \, \hat{t}_3,\quad t_4=\frac{|z_{12}||z_{23}|}{|z_{24}|^2|z_{13}|} \, \hat{t}_4,
\end{align}
\end{subequations}
using the Mellin representation of the exponential\footnote{The Mellin integral is defined as $f(z)=\int_{c-i\infty}^{c+i\infty}\frac{ds}{2\pi i}z^{-s}\phi(s)$ where the integral is over the imaginary axis and $c\in\mathbb{R}$ is such that $a< c < b$ and $\phi(s)$ is analytic in $a< \text{Re}(s)< b$ and $\phi(s)\to 0$ uniformly in the same range as $s\to \pm i \infty$.}
\begin{align}
\exp[-z_{ij}^2]=\int_{c-i\infty}^{c+i\infty}\frac{ds_{ij}}{2\pi i}\,\Gamma(s_{ij})(z_{ij}^2)^{-s_{ij}},
\end{align}
and performing the Gaussian integrals in $\hat{t}_i$ we finally get
\begin{equation}
\label{eq:MellD}
D_{\Delta_1 \Delta_2 \Delta_3 \Delta_4}(z_i)= \frac{\pi^{d/2}\Gamma\left(\frac{\hat{\Delta}-d}{2}\right)}{\Gamma(\Delta_1)\Gamma(\Delta_2)\Gamma(\Delta_3)\Gamma(\Delta_4)}\left[\prod_{i<j}^{4}\int_{c-i\infty}^{c+i\infty}\frac{ds_{ij}}{2\pi i} \, \Gamma(s_{ij})(z_{ij}^2)^{-s_{ij}}\right] ,
\end{equation}
that means that the Mellin transform of the $D_{\Delta_1 \Delta_2 \Delta_3 \Delta_4}(z_i)$ is a constant function 
\begin{equation}
    M(s_{ij})=\frac{\pi^{d/2}\Gamma\left(\frac{\hat{\Delta}-d}{2}\right)}{\Gamma(\Delta_1)\Gamma(\Delta_2)\Gamma(\Delta_3)\Gamma(\Delta_4)} \,,
\end{equation}
in agreement with \cite{Penedones:2010ue}.

Since in the correlators we computed usually appears more generic terms let us define the generic terms inside the correlators $\mathcal{C}_{\Delta_1\Delta_2\Delta_3\Delta_4}(z_i)$ as
\begin{equation}
\label{eq:F}
    \mathcal{G}^{(p,q)}_{\Lambda_1\Lambda_2\Lambda_3\Lambda_4}(u,v)=u^pv^q\hat{D}_{\Delta_1+\Lambda_1 \Delta_2+\Lambda_1 \Delta_3+\Lambda_3 \Delta_4+\Lambda_4}(u,v)  ,
\end{equation}
where $p,q,\Lambda_i\in\mathbb{Z}$ and $u,v$ are the usual cross ratios defined in \eqref{eq:crossratio}.
We define also the following quantities\footnote{We also have the variable $u=-(p_2+p_4)^2$. Beacuse of conservation of momentum we have the relation between the three Mandelstam variables $s+t+u=\Delta_1+\Delta_2+\Delta_3+\Delta_4$.}
\begin{subequations}
\begin{align}
s=-(p_1+p_2)^2=-(p_3+p_4)^2,\quad t=-(p_1+p_4)^2=-(p_2+p_3)^2 \,,
\end{align}
\end{subequations}
and writing the variables $s_{ij}$ in \eqref{eq:sij} in terms of the above Mandelstam variables we get that the correlators can be written in Mellin space as
\begin{subequations}
\label{eq:CM}
\begin{align}
\mathcal{C}_{\Delta_1\Delta_2\Delta_3\Delta_4}(z_i)&=\frac{z_{23}^{\Delta_1-\Delta_2-\Delta_3+\Delta_4}z_{24}^{\Delta_1-\Delta_2+\Delta_3-\Delta_4}}{z_{12}^{2\Delta_1}z_{34}^{\Delta_1-\Delta_2+\Delta_3+\Delta_4}}\int\frac{ds \, dt}{(2\pi i)^2}u^{\frac{s}{2}+\frac{\Delta_1-\Delta_2}{2}}v^{\frac{t}{2}-\frac{\Delta_1+\Delta_4}{2}}M(s,t)\bar{\Gamma}(s,t)  \,,
\end{align}
\end{subequations}
where we have defined the $\Gamma's$ block as
\begin{subequations}
\begin{align}
\bar{\Gamma}(s,t)&\equiv\Gamma\left(-\frac{s}{2}+\frac{\Delta_1+\Delta_2}{2}\right)\Gamma\left(-\frac{s}{2}+\frac{\Delta_3+\Delta_4}{2}\right)\Gamma\left(-\frac{t}{2}+\frac{\Delta_1+\Delta_4}{2}\right)\nonumber\\
&\times\Gamma\left(-\frac{t}{2}+\frac{\Delta_2+\Delta_3}{2}\right)\Gamma\left(\frac{s+t}{2}-\frac{\Delta_1+\Delta_3}{2}\right)\Gamma\left(\frac{s+t}{2}-\frac{\Delta_2+\Delta_4}{2}\right) \,.
\end{align}
\end{subequations}

We focus on the case $d=2$, $\Delta_1=\Delta_2$, $\Delta_3=\Delta_4$ even if the generalization is doable for generic $\Delta_i$ but more involved. The generic term appear inside the correlators as
\begin{equation}
\mathcal{C}_{\Delta_1\Delta_1\Delta_3\Delta_3}(z_i)=\frac{1}{z_{12}^{2\Delta_1}z_{34}^{2\Delta3}}\left[\mathcal{G}^{(p,q)}_{\Lambda_1\Lambda_2\Lambda_3\Lambda_4}(u,v)\,+\cdots\right] .
\end{equation}
The logic to find the Mellin representation of this term inside the correlator is to put it in the same form of \eqref{eq:CM} and considering the function $M(s,t)$ as the unknown. 
Concretely, the steps to follow are firstly noticing that we are dealing with $\hat{D}$-function and so we have to rewrite it as a non-hatted  $D$-function using \eqref{eq:Dbar} and \eqref{eq:relD1}. 
Then we use the Mellin representation of the $D$-function \eqref{eq:MellD}. The factor $u^pv^q$ in front of the D-function and the other factors coming from passing from $\Hat{D}$ to $D$-function will shift the power of the terms $(z_{ij})^{-s_{ij}}$. By performing a change of variable in order to have the correct power as in \eqref{eq:CM} and the same block of $\Gamma$-functions we can extract the function.

The Mellin transform of the generic terms then reads
\begin{subequations}
    \label{eq:MellinF}
    \begin{align}
    M^{(p,q)}_{\Lambda_1\Lambda_2\Lambda_3\Lambda_4}(s,t)=\pi \,  &\mathcal{N}(\Lambda_i,\Delta_i)\left(-\frac{s}{2}+\Delta_1\right)_a\left(-\frac{s}{2}+\Delta_3\right)_b\nonumber\\
    &\times\left(-\frac{t}{2}+\frac{\Delta_1+\Delta_3}{2}\right)_c\left(-\frac{t}{2}+\frac{\Delta_1+\Delta_3}{2}\right)_d\\
    &\times\left(\frac{s+t}{2}-\frac{\Delta_1+\Delta_3}{2}\right)_e\left(\frac{s+t}{2}-\frac{\Delta_1+\Delta_3}{2}\right)_f \,, \nonumber
     \end{align}
\end{subequations}
with
\begin{subequations}
\begin{align}
    a&=p-\Delta_1,\quad b=p+\Delta_1-2\Delta_3+\frac{\Lambda_1+\Lambda_2-\Lambda_3-\Lambda_4}{2}\nonumber\\
    c&=q,\quad\quad\quad d=q-\frac{\Lambda_1-\Lambda_2-\Lambda_3+\Lambda_4}{2}\\
    e&=-p-q+\Delta_3+\Lambda_4,\quad f=-p-q+\Delta_3+\frac{\Lambda_1-\Lambda_2+\Lambda_3+\Lambda_4}{2}    \,, \nonumber 
\end{align}
\end{subequations}
and with the normalization factor given by
\begin{equation}
    \mathcal{N}(\Lambda_i,\Delta_i)=\frac{\Gamma\left(-1+\Delta_1+\Delta_3+\frac{\Lambda_1+\Lambda_2+\Lambda_3+\Lambda_4}{2}\right)}{\Gamma\left(\Delta_1+\Lambda_1\right)\Gamma\left(\Delta_2+\Lambda_2\right)\Gamma\left(\Delta_3+\Lambda_3\right)\Gamma\left(\Delta_4+\Lambda_4\right)} \,,
\end{equation}
and the Pochammer symbol defined as $(x)_a=\frac{\Gamma(x+a)}{\Gamma(a)}$.

\newpage

\providecommand{\href}[2]{#2}\begingroup\raggedright\endgroup

\end{document}